\documentclass[a4paper,11pt]{article}
\pdfoutput=1 
\usepackage{jheppub} 
\usepackage{amsfonts}
\usepackage{textcomp}
\usepackage{amssymb,amsmath}
\usepackage{hyperref}
\usepackage{comment}
\usepackage{graphicx}
\usepackage{multirow,booktabs,multirow}
\usepackage{braket}
\usepackage{bbold}
\usepackage{xcolor}
\usepackage[utf8]{inputenc}
\usepackage{cleveref}
\usepackage{eqparbox}
\usepackage{cancel}
\usepackage{slashed}
\usepackage{units}

\newcommand{\med}[1]{\langle #1\rangle}

\newcommand{\letter}{\emph{Paper I}}

\def \tr {{\rm Tr}}

\def \DM {\pi^0}
\def \Y {\mathcal{Y}}

\title{Pion dark matter in a $\theta$ vacuum: a thermal relic with sharp velocity-dependent self-interactions
 }

\author[a]{Camilo Garc\'ia-Cely,}
\affiliation[a]{Instituto de F\'isica Corpuscular (IFIC), Consejo Superior de Investigaciones
Cient\'ificas (CSIC) and Universitat de Val\`encia,  C/ Catedratico Jose Beltran 2, E-46980 Paterna, Spain}

\author[b]{ Giacomo Landini,}

\author[c]{Luca Marsili,}

\author[d]{ \'Oscar Zapata }
\affiliation[d]{
Instituto de F\'isica, Universidad de Antioquia,\\
Calle 70 \# 52-21, Apartado A\'ereo 1226, Medell\'in, Colombia}

\abstract{As recently proposed, a non-vanishing topological angle may play a central role in QCD-like theories of dark matter (DM). 
In this work, we introduce a dark photon portal to the Standard Model in order to establish thermal equilibrium in the early Universe, and discuss the ensuing  phenomenological constraints, including the stability of DM. %
The resulting dynamics accounts for the observed DM relic abundance and yields velocity-dependent DM self-interactions in astrophysical halos.
Due to the sharp velocity dependence arising from a Breit-Wigner resonance,
dedicated studies are required to assess the gravothermal evolution in detail, especially in the core-collapse regime. This is  particularly timely in light of self-interacting DM interpretations of strong-lensing systems such as SDSS~J0946+1006, which can be naturally explained within our framework.
}


\begin{document}
\maketitle

\section{Introduction}
\label{sec:intro}
Understanding dark matter (DM) remains one of the most pressing open problems in modern science. DM in the form of  dark pions~\cite{Hochberg:2014kqa} constitutes a compelling solution given its similarity with the QCD sector of the Standard Model (SM). In a recent work~\cite{Garcia-Cely:2024ivo}, it was pointed out that the  topological theta angle of the dark sector can substantially modify the dynamics of DM pions. Concretely, Ref.~\cite{Garcia-Cely:2024ivo}, hereafter referred to as \letter, demonstrated that a non-vanishing theta angle induces number-changing processes among the DM pions. As a result, their freeze-out can thermally give rise to the observed DM relic density, particularly in the presence of resonant effects. See also Ref.~\cite{Kamada:2017tsq}, which identified the number-changing processes associated with the $\theta$ angle, but reproduces the relic density with a mechanism independent of $\theta$, 
as well as Refs.~\cite{Hochberg:2015vrg, Kuflik:2015isi,Bernal:2015bla, Bernal:2015xba, Bernal:2015ova, Choi:2015bya, Choi:2016hid,Soni:2016gzf, Kamada:2016ois,  Bernal:2017mqb,Cline:2017tka, Choi:2017mkk,  Kuflik:2017iqs, Heikinheimo:2018esa, Choi:2018iit,Hochberg:2018rjs, Bernal:2019uqr, Choi:2019zeb,  Katz:2020ywn, Smirnov:2020zwf, Xing:2021pkb, Braat:2023fhn, Bernreuther:2023kcg,Garani:2021zrr, Dey:2016qgf,Frumkin:2025iit}  for related work on MeV DM, and Refs.~\cite{Redi:2016kip, Draper:2018tmh, Abe:2024mwa} for the effect of the $\theta$ angle in QCD-like DM models in other contexts.

In this paper, we revisit the framework proposed in Ref.~\cite{Garcia-Cely:2024ivo} and address an aspect that remained unexplored, namely, the interactions between DM and ordinary particles of the SM. A thorough investigation is crucial for the following reasons: on the one hand, such interactions are essential for understanding the thermal processes leading to the relic abundance, which inherently assume that the DM sector thermalizes with the SM sector. On the other hand, with the portal interaction specified, the number-changing processes in the DM sector can lead to indirect detection signatures of DM, potentially including its decay. One must therefore ensure the DM meta-stability.

In contrast to the previous work~\cite{Garcia-Cely:2024ivo}, this analysis is necessarily model-dependent, as multiple realizations are possible~\cite{Hochberg:2015vrg,Hochberg:2018rjs,Kamada:2017tsq,Katz:2020ywn} depending on the portal that communicates  DM and the SM fields. To carry out our study, and for the sake of demonstrating that there exists at least one model that satisfies all current constraints, we focus on the benchmark model BM1 introduced in the \letter. We extend this model by connecting it to the SM via a massive dark photon~\cite{Holdom:1985ag}, endowing the dark sector with an additional $U(1)_d$ gauge interaction.

We examine the Boltzmann equations relevant to determining the DM relic density and justify several assumptions made in the \letter. In fact, we will show that, although correct, such assumptions were overly restrictive and, to a certain extent, unnecessary. This refinement has implications for the allowed theta-angle parameter space and may impact predictions for DM halo structures. For instance, we find a new family of solutions reproducing the observed relic density.

Moreover, the theta angle can also lead to modifications in the structure of small-scale DM halos today~\cite{Spergel:1999mh,Tulin:2017ara}, while still agreeing with the predictions of collisionless dark matter (CDM) on larger scales~\cite{Harvey:2015hha, Bondarenko:2017rfu, Harvey:2018uwf, Sagunski:2020spe, DES:2023bzs}. Thus, the scenario naturally realizes the self-interacting dark matter (SIDM) paradigm~\cite{Spergel:1999mh}. When both effects are taken into account, one obtains a viable model of SIDM wherein the relic density is governed by the same physics responsible for altering DM halo structures. This result is highly non-trivial, as it is often believed that strongly coupled models of DM, while potentially predicting DM self-interactions, cannot simultaneously account for velocity-dependent scattering and therefore cannot capture the size-dependent effects in halo dynamics. Comprehensive reviews can be found in Refs.~\cite{Tulin:2017ara,Adhikari:2022sbh}.

In particular, we now include a discussion of systems such as SDSS\,J0946+1006~\cite{Vegetti:2009cz}, whose properties suggest an SIDM halo undergoing gravothermal collapse, see e.g~\cite{Minor:2020hic}, and which can be naturally explained within our framework through its sharp velocity dependence with enhanced cross sections at intermediate scales and suppressed interactions in clusters.

The structure of this paper is as follows. We begin by reviewing QCD-like theories of DM in Section~\ref{sec:recap}. In Section~\ref{sec:be}, we discuss the Boltzmann equations in full generality, maintaining a model-independent perspective. The portal to the Standard Model is introduced in Section~\ref{sec:portal}, where we examine thermalization, DM stability, and indirect detection constraints.  In Section~\ref{sec:halos}, we discuss the implications in DM halos.
Finally, we present our conclusions in Section~\ref{sec:conclusions}.

\section{Pion DM in a $\theta$ vacuum}
\label{sec:recap}

\subsection{Pion DM for unitary gauge groups}
We consider  QCD-like theories with gauge group $SU(N_c)$ and $N_f$ flavors of light Dirac quarks $q$ in the fundamental representation. Our results can be easily generalized to other gauge group choices such as $SO(N_c)$ and $Sp(N_c)$. Then, the Lagrangian of the theory is given by
\begin{equation}
	\mathcal{L} = -\frac{1}{4}F_{\mu\nu}F^{\mu\nu} + \frac{g^2\theta}{32\pi^2} F_{\mu\nu}\widetilde{F}_{\mu\nu} + \bar{q}i\slashed{D} q - \left(\bar{q}_L M q_R + \text{h.c.}\right),
 \label{eq:L}
\end{equation}
where $M = \text{diag}(m_1,\ldots,m_{N_f})$ is the quark mass matrix with $m_q \neq 0$ and real for all flavors. Strong interactions confine at some energy scale $\Lambda$ and generate a fermion condensate, $\langle\bar{q}q\rangle \sim \Lambda^3$~\cite{Bai:2013xga}, in analogy with standard QCD. If $m_q\ll \Lambda$, this Lagrangian exhibits an approximate $SU(N_f)_L \times SU(N_f)_R$ symmetry, corresponding to independent chiral transformations of the left- and right-handed fermion components in flavor space.
In particular, an axial transformation takes the form
\begin{align}
q_L \rightarrow e^{-i\theta Q/2} q_L\,, &&  \text{and}&&q_R \rightarrow e^{i\theta Q/2} q_R \,,
\end{align}
where $Q$ is an arbitrary $N_f\times N_f$ hermitian matrix. Being anomalous under the $SU(N_c)$ gauge interactions, such a transformation induces not only a change in the mass matrix but also a shift in the $\theta$ angle. Concretely,
\begin{align}
\theta \to \theta(1 - \text{Tr}\,Q)\,, 
&&
\text{and}
&&
M \to M_\theta= e^{i\theta Q/2} M e^{i\theta Q/2}\,.
\label{eq:Mtheta}
\end{align}
Hence, one can effectively move the $\theta$ parameter from the $F\tilde{F}$ term into the mass matrix by imposing 
\begin{equation}
\text{Tr}\,Q = 1   \,.
\label{eq:traceCondition}
\end{equation}
In particular, one may take $Q$ diagonal such that $Q=\text{diag}(q_1,..,q_{N_f}) $. 

It is believed that the fermion condensate spontaneously breaks the chiral symmetry to its diagonal subgroup, $SU(N_f)_L \times SU(N_f)_R \rightarrow SU(N_f)_V$. This leads to the emergence of $N_f^2 - 1$ pseudo-Goldstone bosons, $\pi^a$, whose dynamics at sufficiently low energies is described in chiral perturbation theory~\cite{Pich:1991fq,Scherer:2002tk} by  
\begin{align}\label{eq:chiralLag}
	\mathcal{L}_{\rm eff} = \frac{f_\pi^2}{4} \text{Tr}[\partial_\mu U^\dagger \partial^\mu U] + \frac{f_\pi^2}{2} B_0 \text{Tr}[M_\theta^\dagger U+ U^\dagger M_\theta]\,, &&
    \text{with}
    && U =  e^\frac{i\pi^a\lambda^a}{f_\pi} \,.
\end{align}
We normalize the generators of $SU(N_f)$ such that $\operatorname{tr}[\lambda^a \lambda^b] = 2\delta^{ab}$, and parametrize the fermion condensate in terms of the dark meson constant, $f_\pi$, as $\langle \bar{q} q \rangle = -B_0 f_\pi^2$. Note that it is  expected that the confinement scale is of the order $\Lambda\sim 4\pi f_\pi/\sqrt{N_c}$~\cite{Kamada:2022zwb}.

Let us emphasize a few points. The broken generators of the axial $SU(N_f)_A$ subgroup, corresponding to the chiral transformations, $q_{L,R}\to \exp[\mp i\alpha^a\lambda^a]q_{L,R}$, are in a one-to-one correspondence with the light pseudo-scalar mesons $\pi_a$. Also,
the $SU(N_f)_V$ symmetry is an exact symmetry only in the limit of degenerate quarks; in general, it is explicitly broken by the differing values of the quark masses and is only approximately conserved.

\paragraph{Cancellation of tadpoles.} The vacuum  is unstable unless the linear terms in  $\pi^a$ are absent from $\mathcal{L}_{\rm eff}$. These terms cancel when 
$\text{Tr}[M_\theta \lambda^a -\lambda^a M_\theta^\dagger] =0$, or equivalently, $  \text{Tr}[\sin(\theta Q) M \lambda^a ]=0 $ for every $a$. Since the matrices $\lambda^a$  generate the space of traceless  $N_f\times N_f$ hermitian matrices,   $\sin(\theta Q) M$ must be proportional to the identity. This happens if there exists a function $\alpha(\theta)$ such  that
\begin{align}
\sin(\theta Q)  = \alpha(\theta) \, M^{-1}\,, 
&&\text{or in components}
&&
 m_i  = \frac{\alpha(\theta)}{\sin ( q_i\theta  )}\,.
 \label{eq:sinQcondition}
\end{align}
In particular, one can cast $M_\theta$ in Eq.~\eqref{eq:Mtheta}  as
\begin{equation}
M_\theta = \cos(\theta Q)M  +i \alpha(\theta) \,  1\!\!1\,,
\end{equation}
Likewise, the $M$-dependent part of the effective Lagrangian is 
\begin{eqnarray}
\label{eq:chiLagrangian}
\mathcal{L}_{\rm eff} &\supset& \frac{f_\pi^2 B_0}{2} \left(\tr[\cos(\theta Q) M (U + U^\dagger)] -i \alpha (\theta) \tr[U - U^\dagger]\right)\\
&=&\frac{f_\pi^2 B_0 \alpha(\theta)}{2} \left(\tr[\cot(\theta Q)  (U + U^\dagger)] -i \tr[U - U^\dagger]\right) \nonumber.
\end{eqnarray}
The function $\alpha(\theta)$ can be found imposing the condition in Eq.~\eqref{eq:traceCondition}. One implicit solution\footnote{There are more solutions corresponding to the different branches of the $\arcsin$.} is
\begin{align}
\theta =  \sum_i \arcsin \left({\frac{\alpha(\theta)}{m_i}}\right), && \text{and}&& Q = \frac{1}{\theta} \arcsin\left(\alpha(\theta) M^{-1}\right) \,.
\label{eq:oneQsolution}
\end{align}
Explicit solutions can be found in two important cases:
\begin{itemize}
\item For a quark degenerate mass spectrum, $m_i =m$, for which
\begin{align}
 \alpha(\theta) =m \sin\left(\theta/N_f\right)\,,&&\text{and}&&   Q=\frac{1}{N_f} \,  1\!\!1 \,.
\end{align}
This regime was studied in Ref.~\cite{Kamada:2017tsq}. Since all meson masses are identical in this case, several effects are not captured; in particular,  no resonant effects can arise.

\item For $\theta\ll1$, according to Eq.~\eqref{eq:sinQcondition}, $\alpha(\theta)\ll 1 $ and hence
\begin{align}
 \alpha(\theta) =\frac{\theta}{\tr M^{-1}} \,,&&\text{and}&&   Q=\frac{M^{-1}}{\tr M^{-1}} \,.
 \label{eq:thetaexpansion}
\end{align}
This regime for $\theta$ was assumed in the \letter; however, it was merely a simplifying assumption. As we shall see, allowing for arbitrary $\theta$ does not lead to significant differences.

\end{itemize}
Note that the $ N_f + 1 $   parameters $ \theta $ and $ m_i $ can be exchanged for $ \theta$, $ \alpha$ and $ q_i $ ($i = 1, \ldots, N_f - 1 $) taking them as independent.

\paragraph{Meson mass spectrum.} According to Eq.~\eqref{eq:chiLagrangian}, the meson masses follow from
\begin{equation}
(M_\text{meson}^2)^{ab} =\frac{1}{2}B_0 \alpha(\theta)\tr\left( \cot(\theta Q)\,\{\lambda^a,\lambda^b\}\right) \,,
\end{equation}
with $a,b=1,\dots,N_f$.
In the particular cases for which $|q_i \theta| <\pi/2$,

Eq.~\eqref{eq:sinQcondition} allows to write $\cos(q_i \theta ) = \sqrt{1- (\alpha(\theta)/m_i)^2}  $. This leads to $ \alpha(\theta)\cot(q_i \theta ) = m_i \sqrt{1- (\alpha(\theta)/m_i)^2}  $, and hence
\begin{equation}
(M_\text{meson}^2)^{ab} = \frac{1}{2}B_0 \tr\left( \sqrt{M^2- \alpha(\theta)^2}\,\{\lambda^a,\lambda^b\}\right)  \,.
\end{equation}
Since the function $\alpha(\theta)$ is a number, in such cases, the sole effect of the $\theta$ angle on the meson masses with respect to a theory with $\theta=0$ is to replace the quark masses by 
\begin{equation}
m_i \to m^{\rm eff}_i \equiv \sqrt{m_i^2-\alpha(\theta)^2}\,.
\label{eq:formalRep}
\end{equation}
In particular, assuming $m_1 <m_2 <\ldots< m_{N_f}$, we have   $m^{\rm eff}_1 <m^{\rm eff}_2 <\ldots< m^{\rm eff}_{N_f}$. 
\paragraph{Number-changing interactions. } While the first term in Eq.~(\ref{eq:chiLagrangian}) describes meson masses and interactions involving an even number of mesons, the second term leads to interactions with an odd number of mesons. The leading two terms in the chiral expansion of the latter give 
\begin{equation}
\mathcal{L}_\text{eff, odd terms} = -\frac{B_0 \alpha(\theta)}{3f_\pi\, } \left(d_{abc} \pi_a \pi_b \pi_c - \frac{c_{abcde}}{10f_\pi^2} \pi_a \pi_b \pi_c \pi_d \pi_e \right)+ {\cal O}(f_\pi^{-5})\,.
\label{eq:thetaterms}
\end{equation}
With  $ d_{abc}=\frac{1}{4}\tr(\{\lambda_a,\lambda_b\}\lambda_c)$, and  $c_{abcde}=(\delta_{ab}d_{cde}+\delta_{cd}d_{abe})/N_f+\frac{1}{2}d_{abm}d_{cdn }d_{mne}$, which are non-zero only for $N_f \geq 3$. 
As already mentioned, in \letter\ it was assumed that $\theta \ll 1$, which justifies using the expression for $\alpha(\theta)$ given in Eq.~\eqref{eq:thetaexpansion}. However, formally speaking, the effective Lagrangian with the odd terms in $\pi^a$ remains the same for arbitrary $\theta$. Consequently, the results presented in that work concerning number-changing processes remain largely valid even for large values of $\theta$.

Finally, Eq.~\eqref{eq:thetaterms} must be compared against the  Wess-Zumino-Witten (WZW) term~\cite{Wess:1971yu,Witten:1983tw}, an interaction of topological origin that reads  
 \begin{eqnarray}
 \label{eq:WZW}
 \mathcal{S}_{\mathrm{WZW}}&=& -i \frac{N_c}{240 \pi^2}\int \operatorname{Tr}\left(U^{\dagger} \mathrm{d} U\right)^5 \\
 &=&\frac{ N_c}{240 \pi^2 f_\pi^5}  \int d^4 x \,\epsilon^{\mu \nu \rho \sigma} \operatorname{Tr}\left[\lambda^a \lambda^b \lambda^c \lambda^d\lambda^e \right] \pi^a \partial_\mu \pi^b \partial_\nu \pi^c \partial_\rho \pi^d \partial_\sigma \pi^e + {\cal O}(f_\pi^{-7})\nonumber  \,.
 \end{eqnarray}
As Eq.~\eqref{eq:thetaterms}, it gives rise to 3-to-2 meson annihilations. 

\subsection{Our benchmark model}\label{sec:benchmark}
The general properties outlined above were noted in \letter\ and illustrated using two benchmark models. In this paper, we focus on BM1. Let us briefly summarize this model. It is based on a unitary gauge group with $N_f = 3$ and quark mass matrix $M = \text{diag}(m_1, m_2, m_3)$. The resulting meson spectrum resembles that of the SM, featuring mesons $\pi^0$, $\pi^\pm$, $K^0$, $\bar{K}^0$, $K^\pm$, and $\eta$.\footnote{Throughout this work, we will refer to the dark mesons with this notation; unless specified otherwise, SM mesons will not be referred to in this way.
In particular, the superscript does not refer to any charge.}
The DM candidate is identified with the lightest meson,  $\pi^0$, with the rest of the mesons eventually decaying or annihilating into DM.

\begin{figure}[t!]
    \centering
    \includegraphics[width=0.49\linewidth]{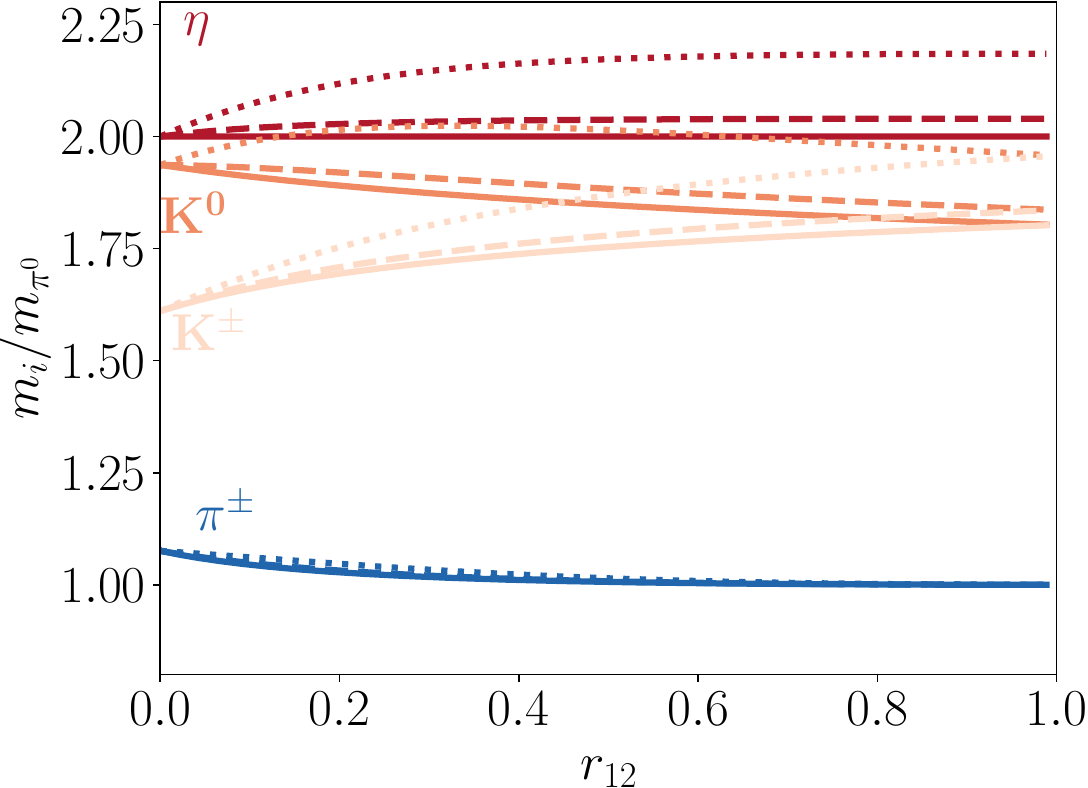}
    \includegraphics[width=0.5\linewidth]{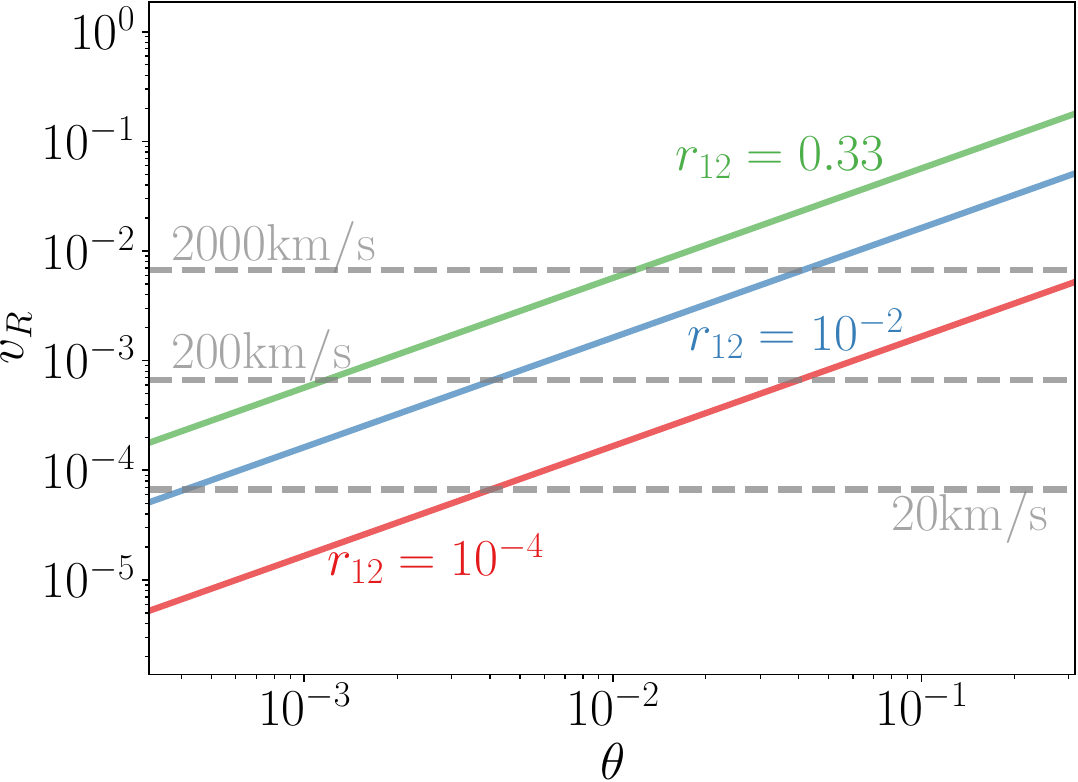}  
    \caption{ {\bf \textit{Left}}: 
    ratios of dark meson masses as a function of $r_{12}$ for fixed values of the vacuum angle: solid, dashed, and dotted lines respectively correspond to $\theta=0, \, \pi/4, \, \pi/2$. {\bf \textit{Right}}: 
    mass combination, $v_R$, defined in Eq.~\eqref{eq:vR} as a function of $\theta$ for fixed values of the indicated quark mass ratio,  $r_{12}$. For reference, the dashed lines correspond to typical velocities in DM halos when expressed in natural units.   
    }  \label{fig:thetavR}  
\end{figure}

\paragraph{Mass spectrum for $\theta=0$. }
In this case, 
\begin{equation}
    \begin{cases}
        m_{\pi^\pm} ^2=B_0(m_1+m_2),\\
        m_{K^\pm}^2=B_0(m_1+m_3),\\
        m_{K,\bar{K}^0}^2=B_0(m_2+m_3),
    \end{cases}
\end{equation}
while $\pi_3$ and $\pi_8$ mix with their masses squared given by the eigenvalues of the matrix
\begin{equation}
  \bigg(
    \begin{array}{cc}
       B_0(m_1+m_2)  &  \frac{B_0}{\sqrt{3}}(m_1-m_2)\\
       \frac{B_0}{\sqrt{3}}(m_1-m_2)  & \frac{B_0}{3}(m_1+m_2+4m_3)
    \end{array}
    \bigg)\,.
\end{equation}
The physical states $\pi^0$ and $\eta$ are given by
\begin{equation}
\bigg(
    \begin{array}{c}
        \pi^0   \\
        \eta
    \end{array}
    \bigg)
    =
    \bigg(
    \begin{array}{cc}
       \cos{\theta_{\eta\pi}}  &  \sin{\theta_{\eta\pi}}\\
       -\sin{\theta_{\eta\pi}}  & \cos{\theta_{\eta\pi}}
    \end{array}
    \bigg)
    \bigg(
    \begin{array}{c}
        \pi_3   \\
        \pi_8
    \end{array}
    \bigg)\,,
\end{equation}
with a mixing angle
\begin{equation}\label{eq:mixing}
    \tan(2\theta_{\eta\pi})=\frac{\sqrt{3}(m_2-m_1)}{2m_3-m_1-m_2} \,.
\end{equation}

Assuming $0 < m_1 \leq m_2 \leq m_3$, the resulting mass spectrum exhibits a well-defined hierarchy: the $\eta$ meson is the heaviest state, while the $\pi^0$ is the lightest. The $\pi^\pm$ mesons are the second lightest states, with the mass splitting $m_{\pi^\pm} - m_{\pi^0}$ vanishing in the limit $m_1 = m_2$. This is illustrated in the left panel of Fig.~\ref{fig:thetavR} (solid line)
for which we assume $m_\eta =2 m_{\pi^0} $ when $\theta=0$. Note that then $v_R|_{\theta=0} =0$ where  

\begin{equation}
v_R = 2 \sqrt{\frac{m_\eta - 2m_{\DM}}{m_{\DM}}},
\label{eq:vR}
\end{equation}
which, as we shall see, characterizes the resonant scatterings of DM~\cite{Chu:2018fzy}. More precisely, the process $\DM\DM\to \DM\DM$ receives a resonant enhancement from the exchange of $\eta$ meson for non-relativistic velocities matching $v_R$.

As in \letter, we remain agnostic of the origin of values of $v_R$ of the order of non-relativistic velocities in natural units, which arises from assuming $v_R|_{\theta=0} =0$ above. 
For examples of model-building in QCD-like models  giving rise to this resonant behavior, see e.g.,  Refs.~\cite{Tsai:2020vpi, Csaki:2022xmu} 
(see also Ref.~\cite{ Lee:2025lko}).
\paragraph{The case of a non-vanishing $\theta$ angle.} As explained above, for $|q_i \theta| < \pi/2$, the meson spectrum can be readily computed from the formulae of $\theta=0$ using Eq.~\eqref{eq:formalRep} .  Moreover, the four parameters $m_1$, $m_2$, $m_3$, and $\theta$ can be expressed as functions of the following four quantities: $\theta$, $\alpha$, the velocity $v_R|_{\theta=0}$, and the mass ratio $r_{12} = m_1/m_2$. Notably, since $\alpha$ is the only parameter out of the latter with mass dimension, any dimensionless combination of the meson masses may depend only on $\theta$, $r_{12}$, $v_R|_{\theta=0}$. Motivated by the resonant effect associated with $\eta$, throughout we will focus on $v_R|_{\theta=0} =0$. In Fig.~\ref{fig:thetavR} (left panel), we show the corresponding mass spectrum of the mesons taking into account the contribution of the $\theta$ angle.   

\begin{figure}
    \centering
    \includegraphics[width=0.49\linewidth]{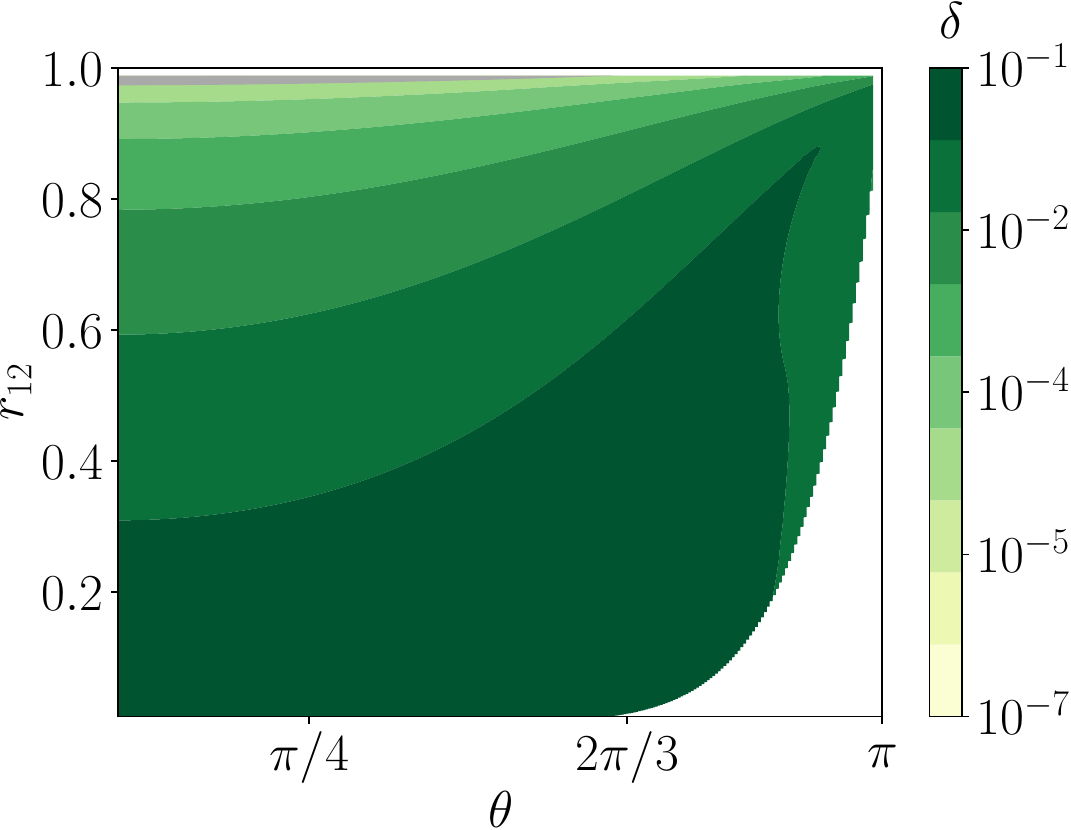}\,\,
    \includegraphics[width=0.49\linewidth]{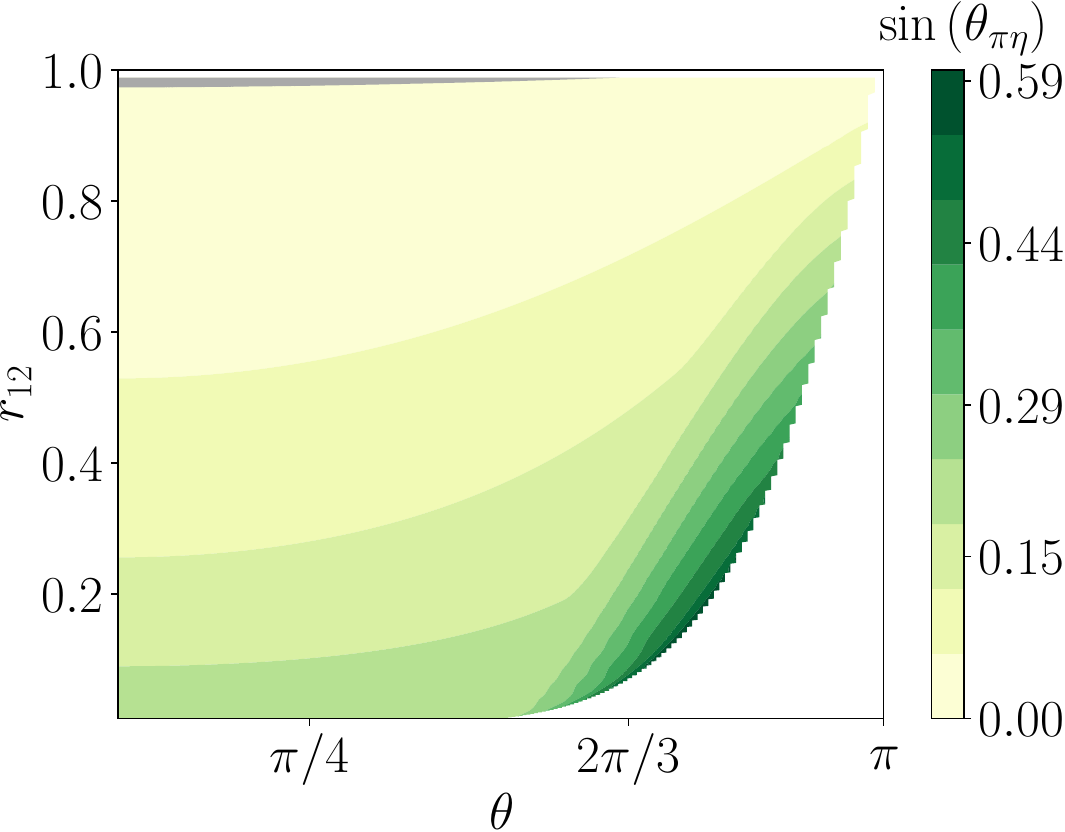}
    \caption{  {\bf \textit{Left}}: Contours of pion mass splitting $\delta$  as a function of $\theta$ and $r_{12}$. The white regions correspond to $|q_i \theta| > \pi/2$. See text for details. {\bf \textit{Right}}: Same as the left panel but for the $\pi^0$-$\eta$ mixing angle $\theta_{\eta \pi}$.}
    \label{fig:r12_theta_contours}
\end{figure}

Taking $v_R|_{\theta=0} =0$, all dimensionless combinations of masses will be functions of $\theta$ and $r_{12}$. 
Three dimensionless combinations of particular importance are the resonance velocity $v_R$, the mass splitting between $\pi^\pm$ and $\pi^0$ parametrized by $\delta \equiv m_{\pi^\pm}/m_{\pi^0} - 1$, as well as the mixing angle $\theta_{\pi\eta}$. They are  respectively shown in the Figs.~\ref{fig:thetavR} (right panel), \ref{fig:r12_theta_contours} (left panel) and  \ref{fig:r12_theta_contours} (right panel). In the white regions of Fig.~\ref{fig:r12_theta_contours} the assumption $|q_i\theta|<\pi/2$ does  not apply.

\paragraph{The role of $\eta$ for number-changing processes and resonant scattering.}
The trilinear interaction in Eq.~(\ref{eq:thetaterms}) --only present for a non-vanishing $\theta$ angle-- induces the decay of $\eta$ meson into two pions as well as resonant elastic scatterings between pions, $\pi^0\pi^0 \to \eta \to \pi^0 \pi^0$. 
For these processes, the rate for $\eta \rightarrow \DM\,\DM$ enters in an important way, and is given by 
\begin{equation}\label{eq:decay}
\Gamma(\eta \rightarrow \DM\,\DM) = \frac{\alpha(\theta)^2 B_0^2 \cos^2(3\theta_{\eta\pi})}{24\pi f_\pi^2 m_\eta } \sqrt{1 - \frac{4m_{\DM}^2}{m_\eta^2}}.
\end{equation}
In addition to this decay and its inverse, the interactions in Eq.~\eqref{eq:thetaterms} induce $3 \leftrightarrow 2$ self-annihilation processes. As such, the $\theta$ angle plays a central role in enabling efficient number-changing dynamics within BM1.

\section{The relic density in the presence of the $\theta$ vacuum}
\label{sec:be}
In our benchmark scenario, as will be explained in section~\ref{sec:halos},   observations of galaxy clusters impose an upper bound on the DM self-scattering cross section of approximately $0.5~\text{cm}^2/\text{g}$~\cite{Harvey:2015hha, Bondarenko:2017rfu, Harvey:2018uwf, Sagunski:2020spe, DES:2023bzs}.
In the region where chiral perturbation theory is valid, namely $m_{\pi^0}/f_\pi < 4\pi/\sqrt{N_c}$, and where this bound is satisfied, the maximum strength allowed for the $3 \to 2$ processes  does not yield sufficient DM annihilation if $v_R \gg 1$, resulting in a relic abundance significantly larger than the observed DM density.  
This issue becomes even more severe in the limit $\theta \to 0$, where the WZW term is the sole source of number-changing interactions, as the resulting $3 \leftrightarrow 2$ processes are velocity suppressed and the scattering in halos is velocity independent.~\cite{Hochberg:2014kqa,Kamada:2022zwb}. In contrast, in the regime $v_R\lesssim 1$, the exchange of the $\eta$ meson resonantly enhances the 3-to-2 processes, reconciling the predicted relic density with the observed value. 
Concretely, the $\eta$ effectively acts as a catalyzer~\cite{Chu:2024rrv}, mediating number-changing interactions such as   $ \DM\,\DM \rightarrow  \eta$ followed by $\eta\,\DM \rightarrow \DM\,\DM$, effectively inducing a resonant 3-to-2 process. 

This section examines how via these number-changing processes a non-zero $\theta$ angle impacts the relic abundances of the mesons.
Before discussing the corresponding Boltzmann equations, let us define notation. 
As usual, 
$H$ is the Hubble rate, which, in a radiation-dominated Universe, is given by 
$H\simeq\sqrt{4\pi^3g_*/45}\,T^2/M_{\rm Pl},
$ with $g_*$ being the energy density degrees of freedom. Likewise, we find it convenient to normalize the number density of mesons to  the entropy density of the Universe
\begin{equation}
Y=\frac{n}{s(T)}\,,
\label{eq:Y}
\end{equation}
where $s(T)=2\pi^2g_{*s}\,T^3/45$,  and $g_{*s}$  is the effective number of degrees of freedom in entropy.

\subsection{The evolution with $ T_{\rm dark}=T$ }

We start by assuming that the dark sector and the SM are in kinetic equilibrium at  the same temperature. 
Later on, we will show that deviations from this assumption have a minor effect on the relic density and do not significantly alter the results for benchmark points of interest. 
Neglecting degeneracy effects, $T_{\rm dark}=T$ means that the dark mesons are distributed in phase space according to 
\begin{equation}
f= \exp\left(\frac{\mu-E}{T}\right) \,.
\label{eq:bdistr}
\end{equation}
The corresponding densities are given by 
\begin{align}
n=\int \frac{d^3 \mathbf{p}}{(2 \pi)^3} f  = \exp\left({\frac{\mu}{T}}\right) n_{{\rm eq}}
&&
\text{with}
&&  n_{{\rm eq}}\equiv \int \frac{d^3 \mathbf{p}}{(2 \pi)^3} \exp\left(-{\frac{E}{T}}\right) \,.
\label{eq:bdistrn}
\end{align}
Hence, the number densities are determined by the chemical potentials, and vice versa, at a given SM temperature. For simplicity, let us first consider a given meson, which could be the DM or a heavier state. Quite generically, the Boltzmann equation determining its number density, when one particle is destroyed in an individual reaction $i\to f$, is given by 
\begin{eqnarray}
\dot{n} + 3 Hn &=&   \left(-\int \left(\prod_{a \in i} \frac{d^3 \mathbf{p}_a}{(2 \pi)^3 2 E_a} f_a  \right)\prod_{b \in f} \left(\frac{d^3 \mathbf{p}_b}{(2 \pi)^3 2 E_b}\right)\left|\mathcal{M}_{i \rightarrow f}\right|^2(2 \pi)^4 \delta^4\left(p_i-p_f\right)\right) \nonumber\\
&&- \left( i \leftrightarrow f \right)\,,
\end{eqnarray}
where the second line accounts for the inverse reaction.  In terms of the yield in Eq.~\eqref{eq:Y}
\begin{align}
\dot{n} + 3 Hn = s H x  \frac{d Y}{dx} \,, &&\text{with} && x=\frac{m_{\pi^0}}{T} \,.
\end{align}

By substituting Eq.~\eqref{eq:bdistr} into the Boltzmann equation, and using the principle of detailed balance --which relates the squared amplitudes of a reaction and its inverse--  we obtain
\begin{eqnarray}
s H x  \frac{d Y}{dx} &=& -\left[\exp\left(\frac{1}{T} \sum_{a\in i}\mu_a\right)-\exp\left(\frac{1}{T} \sum_{b \in f}\mu_b\right) \right] \gamma\left({i \rightarrow f}\right)\nonumber \\
&=& -\left(\prod_{a \in i} \frac{Y_a}{Y_{a,{\rm eq} }}- \prod_{b\in f} \frac{Y_b}{Y_{a,{\rm eq} }}\right) \gamma\left({i \rightarrow f}\right) \,.
\end{eqnarray}
Here,  the interaction rate density is
\begin{equation}
\gamma\left({i \rightarrow f}\right)=\int\left(\prod_{a \in i} \frac{d^3 \mathbf{p}_a}{(2 \pi)^3 2 E_a}e^{-\frac{E_a}{T}} \right)\left(\prod_{b \in f} \frac{d^3 \mathbf{p}_b}{(2 \pi)^3 2 E_b}\right)\left|\mathcal{M}_{i \rightarrow f}\right|^2(2 \pi)^4 \delta^4\left(p_i-p_f\right) \,.
\label{eq:Gamma}
\end{equation}
This quantity is related to the ordinary decay rates and thermally averaged cross sections by
\begin{align}
\begin{cases}
\gamma(1\to23)=n_{1,{\rm eq}}\frac{K_1(m_1/T)}{K_2(m_1/T)}\Gamma(1\to23)\,,
\\
 \gamma(12\to 34) =n_{1,{\rm eq}} \,  n_{2,{\rm eq}}\med{\sigma_{12\to 34} v}\,.
\end{cases}
\end{align}
For two identical particles in the initial state, an additional $1/2$ must be included. The analogue factor for identical particles in the final state is included in the definition of $\Gamma$ and $\med{\sigma v}$.  Adding several reactions and particles is straightforward. 
For simplicity, throughout we  use the  following notation
\begin{align}
   \Y_a =  e^{\mu_a/T} = \frac{Y_a}{Y_{a,{\rm eq}}}.
\end{align}
We are now in a position to write down the Boltzmann equations that determine the DM density, including the effects of coannihilations. Concretely,
the evolution of $\pi_i$ and $\eta$ populations is described by the following coupled Boltzmann equations
\begin{align}\label{eq:beq0}
    sHx\frac{dY_{\pi^0}}{dx} = & -\sum_{i=1,2}\gamma(\eta\pi^0\to\pi_i\pi_i)\left(\Y_{\eta}\,\Y_{\pi^0}-\Y_{\pi_i}^2\right)\nonumber\\
    &+\sum_{i=0,1,2}\gamma(\eta\pi_i\to\pi^0\pi_i)\left(\Y_{\eta}\,\Y_{\pi_i}-\Y_{\pi^0}\,\Y_{\pi_i}\right)\nonumber\\
    &+2\sum_{i=1,2}\gamma(\pi_i\pi_i\to\pi^0\pi^0)\left(\Y_{\pi_i}^2-\Y_{\pi^0}^2\right)+2\gamma_D(\eta\to\pi^0\pi^0)\left(\Y_{\eta}-\Y_{\pi^0}^2\right),\\
\label{eq:beq1}
     sHx\frac{dY_{\pi_i}}{dx} = & +2\gamma(\eta\pi^0\to\pi_i\pi_i)\left(\Y_{\eta}\,\Y_{\pi^0}-\Y_{\pi_i}^2\right) -2\gamma(\pi_i\pi_i\to\pi^0\pi^0)\left(\Y_{\pi_i}^2-\Y_{\pi^0}^2\right)\nonumber\\
     &+2\gamma_D(\eta\to\pi_i\pi_i)\left(\Y_{\eta}-\Y_{\pi_i}^2\right),\hspace{1cm}{\rm for}\,\,i=1,2; \\    
\label{eq:beqeta}
     sHx\frac{dY_{\eta}}{dx} = & -\sum_{i=1,2}\gamma(\eta\pi^0\to\pi_i\pi_i)\left(\Y_{\eta}\,\Y_{\pi^0}-\Y_{\pi_i}^2\right)\nonumber\\
     &-\sum_{i=0,1,2}\gamma(\eta\pi_i\to\pi^0\pi_i)\left(\Y_{\eta}\,\Y_{\pi_i}-\Y_{\pi^0}\,\Y_{\pi_i}\right)\nonumber\\
    & - \sum_{i=0,1,2}\gamma_D(\eta\to\pi_i\pi_i)\left(\Y_{\eta}-\Y_{\pi_i}^2\right).
\end{align}
We neglect the effect of kaons because, although they are lighter than the $\eta$, there are no resonant processes involving them that would  impact the final DM abundance. The remaining contributions are limited to conversion processes, which can be safely neglected since $m_K \gtrsim 1.7\, m_{\pi^0}$.
Furthermore, we also neglect the $3\leftrightarrow2$ self-annihilations originated from both the WZW term and the  5-point interactions induced by $\theta$, as well as  further   number-changing processes, such as 4-to-2  annihilations. This stems from the fact that such processes are subdominant with respect to the resonant $3\to2$ interactions.

The thermally-averaged cross sections entering in the rates are the following. For the semi-annihilations $\eta\,\pi_a\to\pi_b\,\pi_c$ ($a,b,c=0,1,2$)  
\begin{align}\label{eq:cross}
\med{\sigma(\eta\,\pi_a\to \pi_{b}\,\pi_{c}) v}&=\,\frac{\sqrt{5}\,\delta\,m_{\pi^0}^2}{64\,\pi\, f_\pi^4}\,\beta_{abc} +{\cal O}(\theta^2),  
\end{align}
where the nonzero values for $\beta_{abc}$ are 
\begin{align}
    &\beta_{000}=1\,,&&
   \beta_{011}=\beta_{022}=\frac{1}{2}\beta_{0+-}= \frac{529}{81}\,,&&
    \beta_{101}=\beta_{202}=\beta_{\pm 0\pm } =\frac{98}{81}.
\end{align}
For the (co-)annihilations
\begin{align}
    \med{\sigma(\pi_{1,2}\,\pi_{1,2}\to \pi^{0}\,\pi^{0})v}=\med{\sigma(\pi^{+}\pi^{-}\to \pi^{0}\pi^{0})v}=\frac{9\,\sqrt{\delta}\,m_{\pi^0}^2}{32\,\sqrt{2}\,\pi\, f_\pi^4}.
\end{align}

\begin{figure}[t!]
    \centering
    \includegraphics[width=0.59\linewidth]{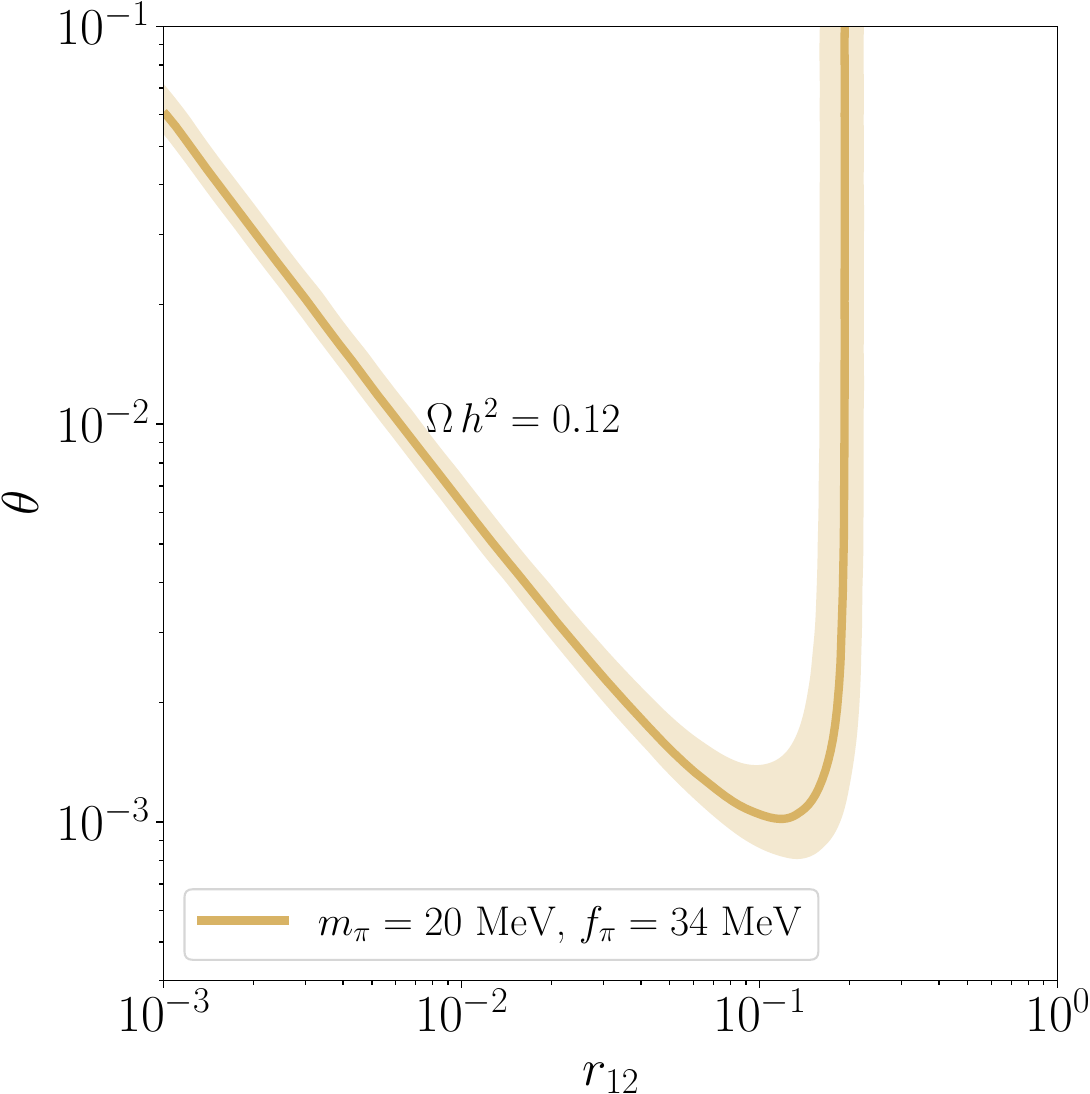}
    \caption{Parameter space leading to  the observed DM relic abundance within $10 \%$~\cite{Planck:2018vyg}  for fixed values of $m_\pi$ and $f_\pi$, assuming kinetic equilibrium with the SM bath. See text for details. 
    }
    \label{fig:thetar12}
\end{figure}

In Fig.~\ref{fig:thetar12}, we present the results of our numerical analysis for a representative parameter choice: $(m_{\pi^0}, f_\pi) = (20~\text{MeV}, 34~\text{MeV})$. We numerically solve the Boltzmann equations~\eqref{eq:beq0}–\eqref{eq:beqeta}, varying $\theta$ and $r_{12}$ within the ranges discussed above with the relic density fixed to the observed value.
Several comments are in order. 

First, we can recognize two  distinct families of solutions: those where the DM abundance depends only on the value of $r_{12}$ and is independent of $\theta$ (on the vertical line), and  those where the abundance is sensitive to both parameters (on the diagonal line). 

\begin{figure}[t!]
    \centering
    \includegraphics[width=0.99\linewidth]{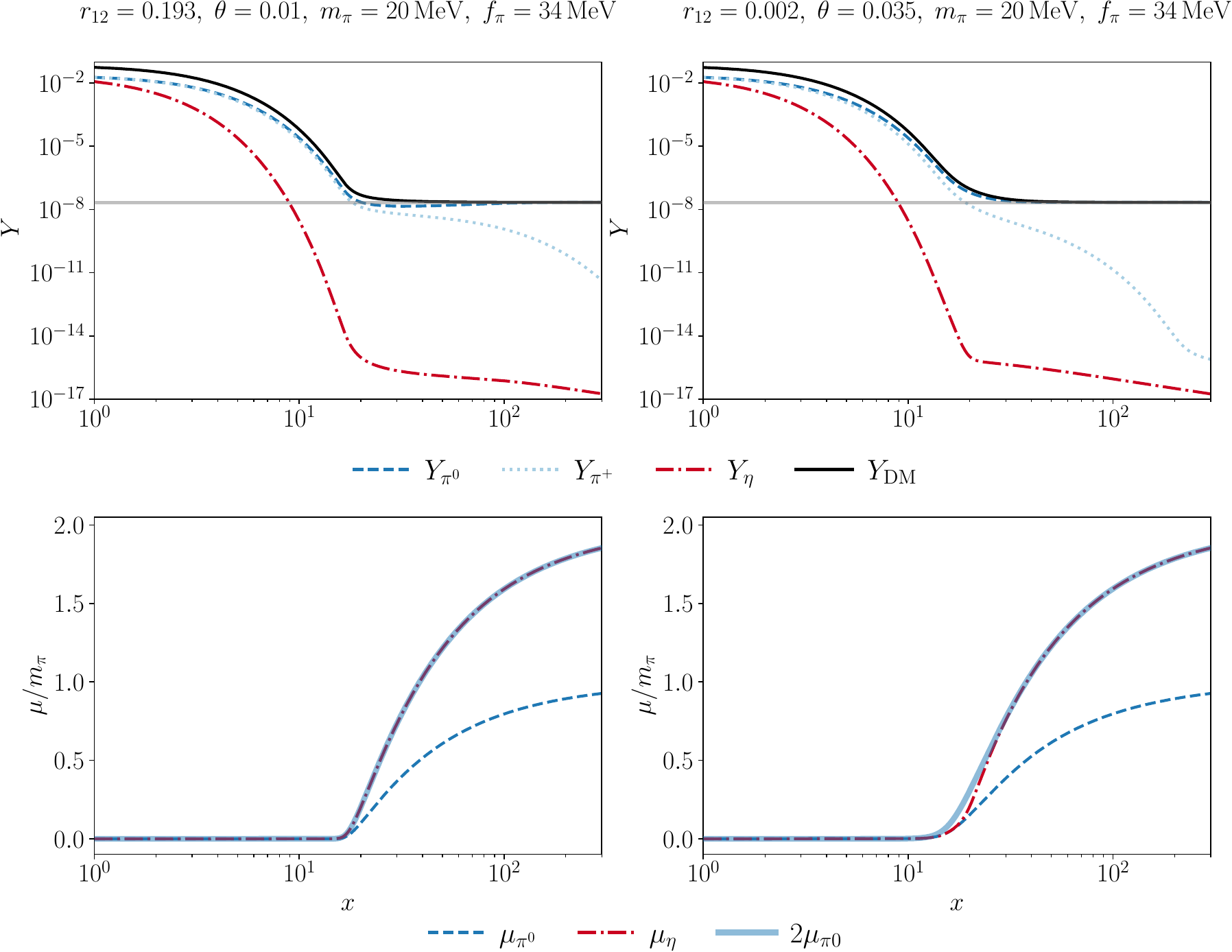}
   \\
   \caption{Dark meson abundances and the respective chemical potentials for two representative points. The point on the left lies on the \textit{vertical line} of Fig.~\ref{fig:thetar12}, while the one on the right belongs to the \textit{diagonal line}. The horizontal line in the upper plots represent the observed DM abundance~\cite{Planck:2018vyg}.
   }
    \label{fig:benchmarkpoints}
\end{figure}

\begin{itemize}
\item 
The former, studied in \letter, are characterized by the fact that  the processes $\eta\leftrightarrow \pi^0\pi^0$ are much faster than the  Universe expansion  when $\eta\pi^0\to\pi^0\pi^0$ freezes out, so that they  keep chemical equilibrium among $\eta$ and $\pi^0$, which enforces $\mu_\eta=2\mu_\pi$  after freeze-out. See also Refs.~\cite{Frumkin:2021zng, Garcia-Cely:2024ivo}. We illustrate this in the left panel of Fig.~\ref{fig:benchmarkpoints} for a representative choice of parameters. Concretely, we show  the evolution of the abundances of $\pi^0,\pi^\pm,\eta$  (upper left panel)  and   their chemical potentials  (lower left panel). Notice that $\mu_\eta = \mu_{\pi^0}=0$ when semi-annihilation and (inverse) decays are simultaneously active, while $\mu_\eta = 2\mu_{\pi^0}$ after the decoupling of semi-annihilations. 
These relations between chemical potentials allow us to write a simplified Boltzmann equation for the combination $Y=Y_\pi+2Y_\eta$:

\begin{equation}\label{eq:BEQfinal1}
    \frac{dY}{dx}=-\med{\sigma_{\eta\pi} v}\frac{sY_{\eta,\rm eq}}{xH}\left(\frac{Y_{\pi}^3}{Y_{{\pi},{\rm eq}}^2}-\frac{Y_{\pi}^2}{Y_{{\pi},{\rm eq}}}\right)\,,
\end{equation}
where $Y_\pi=Y_{\pi^0}+2Y_{\pi^\pm}$ and $\med{\sigma_{\eta\pi}v}$ is the proper combination of cross sections. Notice that, since the mass splitting between $\pi^0$ and $\pi^\pm$ is less than $10\%$, the latter  contribute to the DM relic abundance, converting as $\pi^+\pi^-\to\pi^0\pi^0$ some time after freeze-out. Most importantly, this simplified equation is independent of $\theta$, in agreement with our results in Fig.~\ref{fig:thetar12}. As discussed in \letter, the simplified equation can be solved numerically, or alternatively, an approximate analytical solution of the form $Y_{\pi,\rm eq}(x_{\rm fo})$ can be obtained solving for $x_{\rm fo}$ from $H(x_{\rm fo})\simeq n_{\eta,\rm eq}(x_{\rm fo})\med{\sigma_{\eta\pi}v}$. Both of these approaches are in excellent agreement with
the numerical solution of the full set of Boltzmann equations, thus corroborating the results of \letter.

\item On the other hand, the points on the diagonal lines of  Fig.~\ref{fig:thetar12} correspond to a new family of solutions,  not discussed in \letter. In such points, the inverse decay processes  decouple before semi-annihilations.  In the right panel of Fig.~\ref{fig:benchmarkpoints},  we show  the abundances (upper right panel) and the chemical potentials (lower right panel) of the different species for a representative choice of parameters.
Once again, we  observe that all the chemical potentials vanish when both semi-annihilations  and inverse decays are in equilibrium. When the inverse decays become inefficient, the chemical equilibrium between $\eta$ and $\pi$ is  kept by the semi-annihilation, leading to $\mu_\eta = \mu_{\pi^0}$.
This lasts until semi-annihilation freezes out.
After that, both  processes involved are inactive and their rates are comparable in magnitude,  such that the chemical equilibrium is not achieved.
Eventually, the inverse decay becomes active again, restoring the chemical equilibrium and fixing the chemical potentials to $\mu_\eta = 2 \mu_\pi$.
A similar behavior of the chemical potentials has been identified in other DM scenarios~\cite{Garny:2017rxs, DAgnolo:2017dbv, Puetter:2022ucx}.

In this case, a simplified Boltzmann equation, such as Eq.~\eqref{eq:BEQfinal1}, fails to provide results in quantitative agreement with the full numerical solution. Nevertheless,  we signal that a simplified equation involving only the (inverse) decay processes and enforcing $\mu_\pi=\mu_\eta$ provides results in qualitative agreement with the full numerical solution.

\end{itemize}

\subsection{Departure from kinetic equilibrium}\label{sec:generalT}
Before discussing the thermalization details—and thus the SM portal—let us emphasize that the results so far assume $T = T_{\text{dark}}.$ 
On the one hand, this requires strong interactions for the dark mesons  so that the dark sector develops a thermal distribution. On the other hand, this also requires that the dark sector and the SM are kept in kinetic equilibrium,  occurring if the elastic scatterings among the dark mesons and SM particles are sufficiently strong to allow an efficient energy transfer from one sector to the other.

If kinetic equilibrium between the two sectors is never realized, the dark sector develops its own temperature $T_{\rm dark}$ in view of elastic scattering among dark sector particles, which evolves independently of the SM temperature $T$.
We consider the more predictive scenario in which the interactions between the SM and the dark sectors are strong enough to keep kinetic equilibrium among them initially, thus enforcing $T_{\rm dark}=T$ until the Universe cools down to some temperature, $T_\text{dec}$. At that time, known as kinetic decoupling,  elastic scatterings of the dark mesons off SM particles
become slower than the expansion of the Universe and the dark sector evolves with its own temperature. If number-changing processes within the dark sector are inefficient at the time of kinetic decoupling and afterwards, the temperature evolves as $T_{\rm dark}\sim 1/a$ if DM is relativistic or $T_{\rm dark}\sim 1/a^2$ if DM is non-relativistic~\cite{Pappadopulo:2016pkp,Katz:2020ywn}. See Ref.~\cite{Bernal:2015ova} for a discussion in the context of SIDM.
Notice that in this scenario, the evolution of the dark temperature is predicted in terms of that of the SM.  

In the following, we denote the time of kinetic decoupling as $x_{\rm dec}=m_{\pi^0}/T_{\rm dec}$ and define the time of DM freeze-out as $x_{\rm fo}=m_{\pi^0}/T_{\rm fo}$. In this way, for $x<x_{\rm fo}$ the dark mesons have a vanishing chemical potential enforced by number-changing processes, while chemical decoupling occurs for $x>x_{\rm fo}$ with the dark mesons acquiring a chemical potential. For the family of solutions corresponding to the vertical line of Fig.~\ref{fig:thetar12}, in which semi-annihilations decouple before inverse decays, $x_{\rm fo}$ corresponds to the usual definition of the time at which  $\eta\DM\to\DM\DM$ becomes inefficient.

In the next section, after introducing an explicit SM portal and the interactions between the SM and the dark mesons, we will be interested in the region of the parameter space where kinetic decoupling occurs after DM freeze-out, $x_{\rm dec}>x_{\rm fo}$.
Since $x_{\rm fo}\sim\mathcal{O}(20)$ and $\pi^0$ is the lightest dark particle, the dark sector is non-relativistic at the time of kinetic decoupling, so that for $x>x_{\rm dec}$
its temperature evolves as $T_{\rm dark}=T\,x_{\rm dec}/x\sim 1/a^2$.
We will take this effect into account  by solving the Boltzmann equations for the dark meson abundances in two steps:

\begin{itemize}
    \item for $x<x_{\rm dec}$ we solve the set of coupled Boltzmann equations in Eq.~\eqref{eq:beq0}-\eqref{eq:beqeta};
    \item for $x>x_{\rm dec}$ we solve the same set of equations replacing the dark meson equilibrium distributions as $n_{a,{\rm eq}}(T)\to n_{a,{\rm eq}}(T_{\rm dark})$, so that $Y_{a,{\rm eq}}\to n_{a,{\rm eq}}(T_{\rm dark})/s(T)$, with $T_{\rm dark}=T x_{\rm dec}/x$.
\end{itemize} 
This typically leads to a reduction of the DM relic abundance of a factor of $\mathcal{O}(10-30)\%$.\footnote{In the opposite regime, i.e $x_{\rm dec}<x_{\rm fo}$, number-changing processes are still active after kinetic decoupling enforcing $\mu_i=0$ for all species and heating up the dark sector, whose temperature evolves logarithmically with the scale factor, $T_{\rm dark}\sim 1/\ln a$~\cite{Carlson:1992fn,Pappadopulo:2016pkp}. }


\section{Portal to the SM}
\label{sec:portal}

\subsection{Dark photon portal and DM decays}\label{sec:DPportal}

As mentioned above, we will assume a dark sector initially in  thermal equilibrium with the SM thermal bath. This is a crucial assumption as, in the absence of an explicit portal, number changing processes within the dark sector would modify the evolution of the dark sector temperature and the corresponding DM population.

In this section,   we substantiate this assumption by constructing explicit realizations of BM1 when it is coupled to the SM by means of a dark photon. This portal~\cite{Holdom:1985ag} has been studied in the context of models of strongly interacting massive particle (SIMP) DM where the relic abundance is obtained by 3-to-2 annihilations induced by the WZW term, see e.g. Refs.~\cite{Hochberg:2015vrg} and~\cite{Katz:2020ywn}.  Although other portals can be considered --notably those induced by axion-like particles~\cite{Hochberg:2018rjs,Kamada:2017tsq}, potentially related to the dark theta angle-- we focus on dark photons to demonstrate the existence of one possible realization of QCD-like DM produced by the theta angle, as well as the internal consistency and phenomenological viability of the framework.

For simplicity, in this section we restrict our analysis to small values of the $\theta$ angle, namely $\theta\ll1$. 
As discussed in the previous sections, we do not expect significant differences for larger values, $\theta\sim\mathcal{O}(1)$. We extend the model by introducing a $U(1)_d$ gauge symmetry, associated with a coupling constant $e_d$.
The corresponding dark photon,
 with mass $m_V$ and kinetic mixing $\varepsilon$, is described by  
\begin{equation}
    \mathcal{L}_{\rm DP}=-\frac{1}{4}V_{\mu\nu}V^{\mu\nu}+\frac{1}{2}m_V^2V_\mu V^\mu+\frac{\varepsilon}{2}V_{\mu\nu}B^{\mu\nu},
\end{equation}
where $B_{\mu\nu}$ is associated with the SM hypercharge. 
We remain agnostic regarding the origin of the dark photon mass, which could arise either through the St\"uckelberg mechanism~\cite{Stueckelberg:1938hvi} or via interactions with a dark Higgs field~\cite{Englert:1964et, Higgs:1964pj}.

On the one hand, as usual, the SM sector carries no dark charge and communicates with the DM only through the parameter $\varepsilon$, that induces a mixing between the dark photon and the SM photon. After diagonalization, the SM photon only couples to the SM electromagnetic current $J_\mu^{\rm SM}$ as $e A^\mu J_\mu^{\rm SM}$, while the physical dark photon couples to both the dark current $J_\mu^{\rm dark}$ and the SM current with coupling $V^\mu(e_d \, J_\mu^{\rm dark}+e\,\varepsilon\,c_W \,J_\mu^{\rm SM})$, where $c_W$ denotes the cosine of the weak mixing angle. Since we are primarily interested in the MeV scale,  we ignore the small mixing between the dark photon and the $Z$ gauge boson,  which is suppressed by $\varepsilon\,(m_V/m_Z)^2$~\cite{Babu:1997st}.
 
On the other hand, the dark quarks are charged under the $U(1)_d$ gauge group, with charges $Q_d={\rm diag}\,(\tilde{q}_1,\tilde{q}_2,\tilde{q}_3)$. The chiral Lagrangian in Eq.~\eqref{eq:chiralLag}  in presence of the dark photon is extended to ~\cite{Scherer:2002tk, Katz:2020ywn, Braat:2023fhn} 
\begin{equation}\label{eq:chiralLA}
\begin{aligned}
\mathcal{L}_{\rm eff} & =\frac{f_\pi^2}{4} \operatorname{Tr}\left[\left(D_\mu U\right)^{\dagger} D^\mu U\right]+\frac{B_0 f_\pi^2}{2} \operatorname{Tr}\left(M_\theta^{\dagger} U+U^{\dagger} M_\theta\right)+c f_\pi^4 e_d^2 \operatorname{Tr}\left(U^{\dagger} Q_d U {Q}_d\right) \\
& -\frac{i e_d^2 N_c}{48 \pi^2} \epsilon^{\mu \nu \alpha \beta} V_{\mu \nu} V_\alpha \operatorname{Tr}\left({Q}_d^2 \partial_\beta U U^{\dagger}-Q_d^2 \partial_\beta U^{\dagger} U-\frac{1}{2} Q_dU Q_d\partial_\beta U^{\dagger}+\frac{1}{2} Q_dU^{\dagger} Q_d\partial_\beta U\right) \\
& +\frac{e_d N_c}{48 \pi^2} \epsilon^{\mu \nu \rho \sigma} V_\mu \operatorname{Tr}\left(Q_d\partial_\nu U U^{\dagger} \partial_\rho U U^{\dagger} \partial_\sigma U U^{\dagger}+Q_dU^{\dagger} \partial_\nu U U^{\dagger} \partial_\rho U U^{\dagger} \partial_\sigma U\right) \\
& + {\cal L}_{\rm WZW}
+ \text{higher order operators}\,.
\end{aligned}
\end{equation}
Here, the covariant derivative is defined as $D_\mu U =\partial_\mu U-ie_d V_\mu\,[Q_d,U]$. The last term of the first line is responsible for the mass corrections of the dark mesons charged under $U(1)_d$, with  $c$ being a constant. Assuming a heavy dark photon, $m_V>\Lambda$, it scales as 
$c\sim(\Lambda/m_V)^2\ln(m_V^2/\Lambda^2)$~\cite{Katz:2020ywn}.

In the presence of the portal, DM stability is no longer guaranteed, as the number-changing processes might induce its decay. Note that this does not hold if a conserved symmetry exists in the dark sector, but this is not the case for BM1, in which the flavor symmetry is broken by quark mass differences, $m_1 \neq m_2 \neq m_3$.
Thus, the DM particle is meta-stable, and its interactions with SM particles determine whether it is sufficiently long-lived.

DM decay is induced by the operator $\pi^0\, V_{\mu\nu}\tilde{V}^{\mu\nu}$ and, in presence of a non-vanishing $\pi^0-\eta$ mixing, by $\eta\, V_{\mu\nu}\tilde{V}^{\mu\nu}$.
For a generic choice of dark quark charges, both operators are generated via the axial anomaly: 
the approximate (spontaneously broken) $SU(3)_A$ global symmetry of the Lagrangian is anomalous under the dark $U(1)_d$ interactions.\footnote{More precisely,  even in the limit in which the dark quarks are massless,  a subset of the currents associated with the $SU(3)_A$ transformations are not conserved because of the quantum anomaly, namely  $\partial^\mu J_{\mu,A}^{3,8}\propto e_d^2 V_{\mu\nu}\tilde{V}^{\mu\nu}{\rm Tr}[\lambda^{3,8}Q_d^2]$ with  $J_{\mu,A}^{3,8}=\bar{q}\,\gamma_\mu\gamma_5\lambda^{3,8}\,q$.} Note that this closely resembles the anomalous decay of neutral mesons into two photons observed in the SM. The presence of the anomaly is encoded in the second line of Eq.~\eqref{eq:chiralLA}, which provides interaction terms proportional to  $(e_d^2/f_\pi)\,\pi_{3,8}\,V_{\mu\nu}\tilde{V}^{\mu\nu}\tr[\lambda^{3,8}Q_d^2]$. These, once expressed in terms of the mass eigenstates, induce the aforementioned operators with couplings suppressed by only one power of $f_\pi$. 
\begin{figure}[t!]
    \centering
    \includegraphics[width=0.35\linewidth]{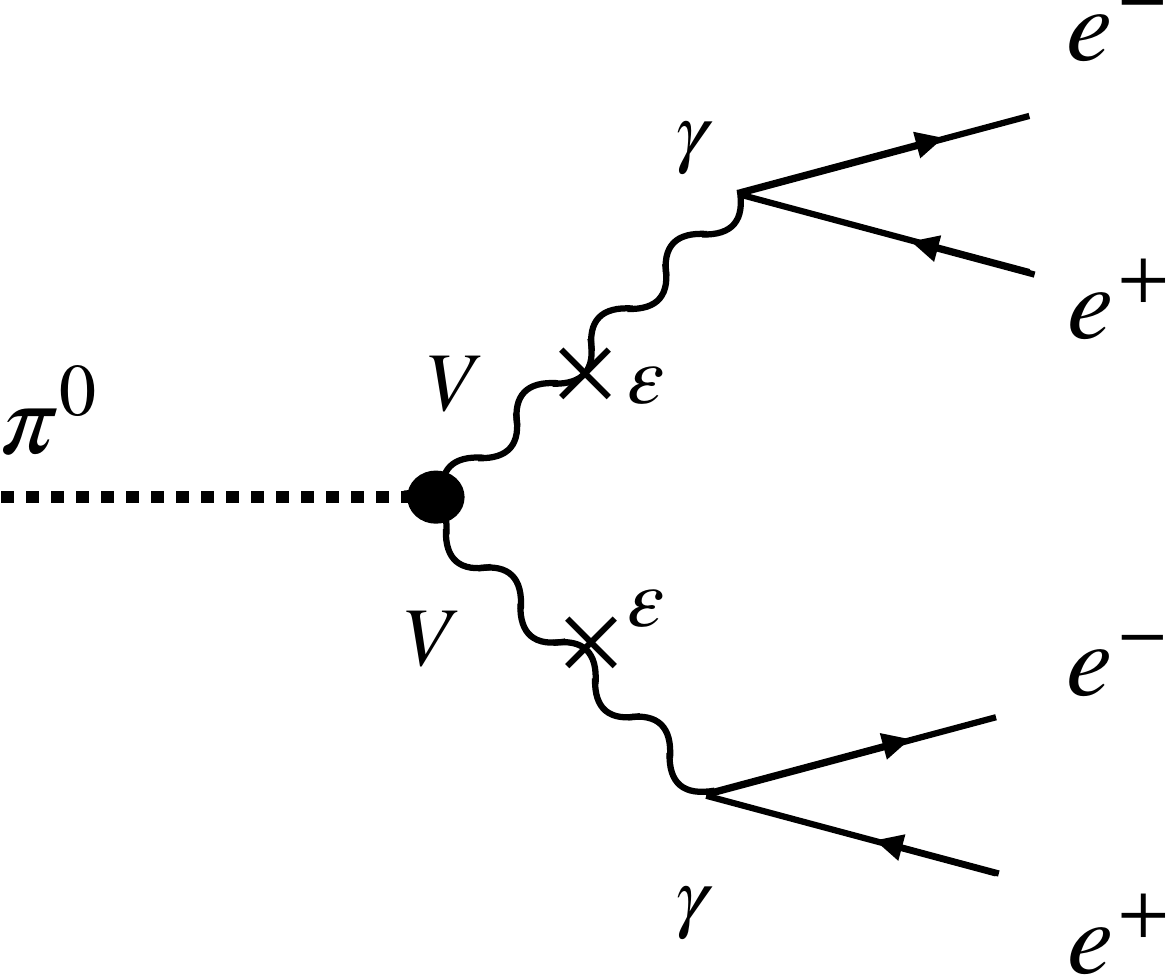}
    \caption{Feynman diagram for the main decay channel of $\pi^0$ in the mass range of interest in this work. A similar diagram holds for the decay of $\eta$.
   }
    \label{fig:diagdecay}
\end{figure}

However, it is possible to choose the charge matrix $Q_d$ in such a way that the anomaly of the axial currents vanishes, ${\rm Tr}[\lambda_{3,8} Q_d^2] = 0$. In such a case, the corresponding currents are conserved in the massless limit, and the anomalous contribution to the operators discussed in the previous paragraph vanishes. This requirement can only be fulfilled by assigning the dark charges so that $\tilde{q}_1^2 = \tilde{q}_2^2 = \tilde{q}_3^2$, and hence $Q_d \propto (\pm1, \pm1, \pm1)$. Without loss of generality, there are only three charge  assignments (up to a global normalization). Two of them are listed in Table~\ref{tab:charge_assignments}, 
 while the third one,  $Q_d^{(3)}\propto{\rm diag}(+1,-1,+1)$, is qualitatively similar to $Q_d^{(2)}$ and provides analogous results.

Even for a vanishing anomaly, the interaction terms $\pi^0 V_{\mu\nu}\tilde{
V}^{\mu\nu}$ and $\eta V_{\mu\nu}\tilde{
V}^{\mu\nu}$ are generated by higher-order operators in the chiral Lagrangian expansion. More precisely, the relevant operators arise at $\mathcal{O}(p^6)$, that is, next-to-next-to-leading order. Following the approach in Ref.~\cite{Katz:2020ywn}, as a representative operator we consider
\begin{equation}\label{eq:Op6}
    \frac{\alpha_d}{(4\pi)^2f_\pi}i\epsilon_{\mu\nu\rho\sigma}V^{\mu\nu}V^{\rho\sigma}\tr(Q_d)\tr(Q_d\,M\,U^\dagger)+{\rm h.c.}\,,
\end{equation}
where $\alpha_d=e_d^2/4\pi$.
This operator induces the $\pi^0 V_{\mu\nu}\tilde{
V}^{\mu\nu}$ and $\eta V_{\mu\nu}\tilde{
V}^{\mu\nu}$ interactions with couplings suppressed by an extra factor of $(m_{\pi^0}/ f_\pi)^2$ compared to the anomalous contribution discussed before.\footnote{Notice that the presence of a non-vanishing $\theta$ angle may induce additional CP-violating operators in the chiral Lagrangian, potentially giving rise to $\theta\,\eta\,V_{\mu\nu}V^{\mu\nu}$ and $\theta\,\pi^0\,V_{\mu\nu}V^{\mu\nu}$ interactions. Their contribution to the DM decay rate is nevertheless suppressed for small values of $\theta$. One could expect $\mathcal{O}(1)$ corrections for large $\theta\sim\mathcal{O}(1)$.} Thus, in the case of vanishing anomaly, the DM decay rate is significantly smaller close to  the chiral limit, $m_{\pi^0}\ll f_\pi$, providing a much larger DM lifetime. 

Finally,  DM is absolutely stable in the limit in which the global $SU(2)_V$ symmetry involving the first two flavors of dark quarks is conserved, i.e. dark isospin. This requires both $r_{12}=0$ and $\tilde{q}_1=\tilde{q}_2$. 
For the cases in which $\tilde{q}_1=\tilde{q}_2$ but $r_{12}\neq0$, the decay of $\pi^0$ is induced by dark isospin breaking, or, in other words, by the $\pi^0-\eta$ mixing, as given in Eq.~\eqref{eq:mixing}. As shown in Fig.~\ref{fig:r12_theta_contours} (right panel), this is fixed by $r_{12}$ and $\theta$.

In the following, we assume $m_V>m_\eta$ so that the decays $\pi^0,\eta\to VV$ are kinematically forbidden. As the dark mesons do not couple to the physical SM photon, the dark decays $\pi^0,\eta\to\gamma\gamma$ are absent and DM decays to SM leptons. For $m_\eta\lesssim 200$ MeV, the relevant DM decay channel is $\pi^0\to e^+e^-e^+e^-$, corresponding to a tree level diagram with two dark photon propagators, see Fig.~\ref{fig:diagdecay}. This process involves the electron-dark photon coupling twice, so that both DM and $\eta$ decay rates are suppressed by $\varepsilon^4e_d^4$. An analogous decay mode arises for the $\eta$ meson. However, its branching ratio is extremely suppressed, as $\eta$ mostly decays to $2\pi^0$ via the $\theta$ angle. We will briefly comment on this below.

\begin{table}[t!]
\centering
\begin{tabular}{ccl}
 \textbf{Charge Assignment} & \textit{\textbf{ P}$^\pm$ } & \textbf{DM Lifetime} 
\\\hline\hline
 $Q_d^{(1)}=\text{diag}(+1,+1,-1)$ &  $K^\pm$  &$\tau_{\rm DM}$ in Eq.~\eqref{eq:DMlifetime} 
 \\
 $Q_d^{(2)}=\text{diag}(+1,-1,-1)$ & $\pi^\pm$ & $\tau_{\rm DM}$ in Eq.~\eqref{eq:DMlifetime} times  $\sin^2\theta_{\pi\eta}$ 
 \\
\end{tabular}
\caption{Charge assignments and associated properties for benchmark models. The two-step thermalization between the DM and the SM bath is achieved by means of scatterings with the meson $P^\pm$. See text for details.}
\label{tab:charge_assignments}
\end{table}

In the following, we will describe the possible charge assignments presented in Table~\ref{tab:charge_assignments} for which the axial anomaly vanishes and the DM  is long-lived.
\begin{itemize}
\item \textit{Charge assignment 1.}  The dark mesons $\pi^0$, $\pi^\pm$ and $\eta$ are neutral under $U(1)_d$, whereas $K^+$ and $K^0$ carry a charge of $+2$, while $K^-$ and $\overline{K^0}$ carry a charge of $-2$. Being charged, the four kaons are stable. Furthermore, the resulting Lagrangian is invariant under the accidental global symmetry generated by $Q'={\rm diag}\,(1,-1,0)$. 
Being the lightest particles charged under this symmetry, the $\pi^\pm$ are stable. In contrast, $\eta$ and $\pi^0$ are not protected by any symmetry, and therefore decay. 
DM decays are induced via the operator in Eq.~\eqref{eq:Op6}, with a lifetime  estimated up to $\mathcal{O}(1)$ factors as 
\begin{eqnarray}\label{eq:DMlifetime}
\tau_{\rm DM}&\sim&  1.4\times 10^{29}\mathrm{\,s\,}\left(\frac{0.1}{\sin\theta_{\pi\eta}}\right)^{2}\left(\frac{10^{-3}}{\alpha_d}\right)^{2}\left(\frac{5 \times10^{-4}}{\varepsilon}\right)^4\nonumber\\
&&\times\left(\frac{m_V} {0.5\,\text{GeV}}\right)^{8}\left(\frac{20\text{ MeV}}{m_{\DM}}\right)^{9}\left(\frac{0.5}{m_{\DM}/f_\pi}\right)^{6}\,.  
\end{eqnarray}
For reference, as shown in Fig.~\ref{fig:r12_theta_contours} (right panel), the range of values for the mixing angle is $0.23\gtrsim\sin\theta_{\pi\eta}\gtrsim 0.005$ for $\theta\ll1$. 
The decay is strongly suppressed close to the chiral limit $m_{\pi^0}\ll f_\pi$, as expected. Furthermore, the DM is stable in the limit $\sin\theta_{\pi\eta}=0$ (equivalent to $r_{12}=0)$ in view of the isospin-preserving charge assignement, $\tilde{q}_1=\tilde{q}_2=1$.
The corresponding constraints from DM indirect searches will be discussed in Section~\ref{sec:constraints}. 

 Moreover, the kaon masses receive a small correction due to the interaction with the dark photon (similarly to the SM).  As long as $m_V>\Lambda$, this can be estimated as $m_K^2\to m_K^2+\Delta m_K^2$ with $\Delta m_K^2\sim 8\,c\, e_d^2\,f_\pi^2$. 
Recall, nevertheless, that the DM relic density is only marginally affected by this.

\item \textit{Charge assignment 2.} The charge of the dark mesons is analogous to the ordinary electric charge in the visible sector: the mesons $\pi^0$, $\eta$, $K^0$, and $\overline{K^0}$ are uncharged, whereas $\pi^\pm$ and $K^\pm$ carry charge $\pm 2$.
While the $\pi^\pm$ states are stable as a result of this,  $\pi^0$ decays with a lifetime estimated --up to factors of $\mathcal{O}(1)$-- using Eq.~\eqref{eq:DMlifetime} without the $(1/\sin^2\theta_{\pi\eta})$ factor. In this case, as $\tilde{q}_1\neq \tilde{q}_2$, the charge matrix explicitly breaks isospin, so that the operator in Eq.~\eqref{eq:Op6} directly induces the decay of $\pi^0$ regardless of its mixing with $\eta$. For the same parameters used in Eq.~\eqref{eq:DMlifetime}, the lifetime is shorter, namely, $\tau_{\rm DM}\sim 10^{27}$~s.

Along with this, the ${\pi^\pm}$ interaction with dark photon shift their mass squared:  $m_{\pi^\pm}^2\to m_{\pi^\pm}^2+\Delta m_{\pi^\pm}^2 $. We can express this in terms of the pion splitting defined in the previous sections as
\begin{equation}
    \delta=\frac{m_{\pi^\pm}-m_{\pi^0}}{m_{\pi^0}}\bigg|_{Q_d^{(2)}}=\delta(r_{12})\bigg|_{e_d=0}+\delta_{V},
\end{equation}
where $\delta(r_{12})\big|_{e_d=0}$ is the correction in the absence of the portal, shown in Fig.~\ref{fig:r12_theta_contours} (left panel).  The other term can be estimated for $m_V>\Lambda$ as~\cite{Katz:2020ywn}

\begin{equation}\label{eq:deltaV}
    \delta_{V}\simeq \frac{\Delta m_{\pi^{\pm}}^2}{2m^2_{\pi^0}}\simeq\frac{\Lambda^2}{m_V^2}\ln\left(\frac{m_V^2}{\Lambda^2}\right)\frac{4\,e_d^2 f_\pi^2}{m_{\pi^0}^2}\,,
\end{equation}
where we used $\Delta m_{\pi^\pm}^2\sim 8ce_d^2f_\pi^2$. 

Notice that the additional source of mass splitting for the charged pions  affects DM relic density when $\delta_V\gtrsim 0.01$. Indeed, the charged pions become sufficiently heavy that their number density is significantly Boltzmann-suppressed at freeze-out, $n_{\pi^\pm} \ll n_{\pi^0}$, thereby reducing the total DM abundance,  $n_{\rm DM} = n_{\pi^0} + 2n_{\pi^\pm}$. Such a reduction in the DM number density requires a larger value of $f_\pi$ to match the cosmological relic abundance for fixed DM mass and $r_{12}$ compared to the case of $\delta_V=0$. Furthermore, if $\delta\gtrsim0.5$, the total cross section for $\eta\pi\to\pi\pi$ becomes smaller, as the $\eta\pi^0\to\pi^+\pi^-$ becomes kinematically closed. 
\end{itemize}
We will take these effects into account in the following section while solving the Boltzmann equations for the dark meson abundances for a representative benchmark point.

\subsection{Thermalization}
In the rest of this section, we discuss the conditions for kinetic equilibrium between the SM and the dark sector. Kinetic equilibrium is established via elastic scatterings of the dark mesons with SM fermions mediated by the dark photon.
Although this discussion shares some similarities with previous studies of other SIMP candidates (see e.g. Refs.~\cite{Hochberg:2015vrg,Katz:2020ywn}), our DM candidate is neutral under $U(1)_d$ and therefore does not interact directly with SM fermions. Thus, kinetic equilibrium is established in two steps:
\begin{itemize}
    \item[\emph{I}.] The lightest dark mesons charged under $U(1)_d$ elastically scatter off SM leptons (mainly electrons and positrons) via the $t$-channel exchange of a dark photon. If these interactions are sufficiently strong, the charged mesons remain in kinetic equilibrium with the SM bath, thereby maintaining a common temperature. The lightest charged mesons are the kaons $K^\pm$ within the charge assignment 1 and the pions $\pi^\pm$ within charge assignment 2, see Table~\ref{tab:charge_assignments}.
The relevant elastic cross section  is given by
\begin{align}
\!\!\!\!\!\!\!
 \sigma v (P^\pm f \to P^\pm  f )  = \frac{e^2 { e_d^2c_W^2 Q_P^2} \, \varepsilon^2 p_f^2}{2 \pi m_V^4},
 &\text{ with } 
 P=\begin{cases}
    K,  \text{ for charge assignment 1}; \\
    \pi,  \text{ for charge assignment 2}
    \,,
\end{cases}
\end{align}
where $p_f$ is the momentum of the SM fermion $f=e^\pm$ and $Q_K=Q_\pi=2$ is the dark charge of the corresponding meson.
They thermalize with the SM 
as long as~\cite{Hochberg:2015vrg,Katz:2020ywn}
\begin{equation}\label{eq:thermalization1}
    \frac{5 \zeta{(5)}}{4} {\frac{T}{m_{P^\pm}}}\Gamma_{\rm scatt} \gtrsim H\,,
\end{equation}
 where the scattering rate is given by
\begin{equation}\label{eq:therm}
\Gamma_{\rm scatt} =\frac{\sum_{P^\pm} Q_P^2}{2}\sum_{f=e^-,e^+}\med{ n_f^{\rm eq}\sigma v (P^\pm\,f \to P^\pm\, f )\, }= \frac{90 \zeta(5) e^2 {e_d^2c_W^2}\varepsilon^2 T^5  }{\pi^3 m_V^4}\,.
\end{equation}
\item[\emph{II}.] The DM particle $\pi^0$ thermalizes with the lightest meson charged under $U(1)_d$, $P^\pm$, thus sharing their temperature, which is the same of the SM bath if the condition in Eq.~\eqref{eq:thermalization1} is satisfied. 
The relevant thermal cross sections are given by
\begin{align}
    \med{\sigma(K^\pm \pi^0 \to K^\pm \pi^0) v}  &=  \frac{m_{\pi^0}^2}{9216 \pi\, f_\pi^4} \frac{\left(13-4 m_{K^\pm}^2/m_{\pi^0}^2\right)^2}{\left(1+m_{K^\pm}/m_{\pi^0} \right)^{3/2}}\sqrt{\frac{8\,m_{\pi^0}}{\pi\,x \, m_{K^\pm} }} \,,\\
    \med{\sigma(\pi^\pm \pi^0 \to \pi^\pm \pi^0) v}  &=  \frac{m_{\pi^0}^2}{1296 \pi\, f_\pi^4} \frac{\left(13-4 m_{\pi^\pm}^2/m_{\pi^0}^2\right)^2}{\left(1+m_{\pi^\pm}/m_{\pi^0} \right)^{3/2}}\left(\frac{m_{\pi^\pm}}{m_{\pi^0}}\right)^4\sqrt{\frac{8\,m_{\pi^0}}{\pi\,x \, m_{\pi^\pm} }} \,.
\end{align}
We numerically find that these expressions are of the same order of magnitude.
Once established, the kinetic equilibrium is maintained 
as long as the elastic scatterings are more rapid than the Hubble expansion, that is, 
\begin{equation}\label{eq:KinEq}
    n_{P^\pm}(T)  \langle \sigma v(P^\pm \pi^0 \to P^\pm \pi^0) \rangle \frac{T}{m_{\pi^0}} \gtrsim H.
\end{equation}
\end{itemize}

The main difference between the two charge assignments is that the charged pions are nearly degenerate in mass with the neutral pion, i.e., $m_{\pi^{\pm}} \sim m_{\pi^0}$, resulting in a significantly larger abundance of charged pions at freeze-out compared to that of kaons, $n_{\pi^{\pm}} \gg n_K$ at $T=T_{\rm fo}$. Consequently, the thermalization condition is more easily satisfied for the charge assignment 2. Thus, for simplicity in the numerical analysis that follows, we choose benchmark points within the charge assignment 2.

Finally, we compute the largest temperatures at which Eq.~\eqref{eq:thermalization1} or Eq.~\eqref{eq:KinEq} no longer holds, which corresponds to the kinetic decoupling $T_{\rm dec}$ as defined in Section~\ref{sec:generalT}.For concreteness, we will focus on parameter space where kinetic decoupling occurs after DM freeze-out, that is $T_{\rm dec}<T_{\rm fo}$.

\subsection{Viable parameter space}

\begin{figure}[t!]
    \centering
    \includegraphics[width=0.80\linewidth]{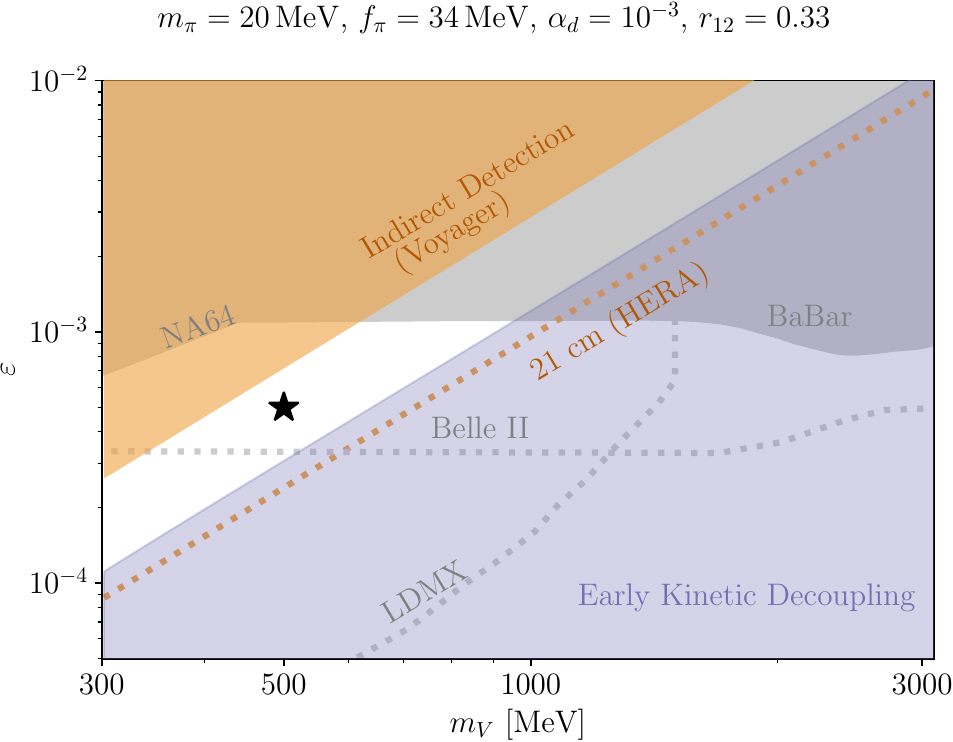}

    \caption{Summary of constraints for the representative choice of the parameters indicated in the legend. 
    In the shaded region at the bottom (light purple), kinetic decoupling occurs before DM freeze-out. The shaded regions in the upper part of the plot represent areas excluded by searches of DM decay (yellow) as well as dark photon searches in collider  and beam dump experiments (gray). Future sensitivities are indicated by the dotted lines: LDMX and Belle II  are shown in gray, while 21 cm prospects are indicated in orange. 
    }
    \label{fig:Therm}
\end{figure}

Taking charge assignment 2,  Fig.~\ref{fig:Therm} shows the region of the parameter space $(m_V,\epsilon)$ where thermalization is achieved for a representative choice of the parameters $m_{\pi^0}$, $f_\pi$, $r_{12}$ and $\alpha_d$. 
In the purple region, denoted as \emph{early thermal decoupling}, at least one of Eqs.~\eqref{eq:thermalization1} and \eqref{eq:KinEq} is violated  before DM freeze-out and kinetic equilibrium is lost. The yellow and gray regions are excluded by indirect detection and collider searches to be discussed below.
Here, the annihilation channel $\pi^+\pi^- \to e^+ e^-$ is slower than the Universe expansion rate at $T=T_{\rm fo}$ and is  irrelevant for the evolution of the DM density\footnote{The corresponding annihilation rate is given by 
\begin{equation}
    \Gamma_{\rm ann}  =\frac{\sum_\pi Q_\pi^2}{N^2_\pi}\med{\sigma v (\pi^+\, \pi^-\ \to e^+e^- )\, n_\pi}= \frac{2 \, e^2 e_d^2c_W^2 m_\pi \varepsilon^2 T}{ \pi (m_V^2-4m_\pi^2)^2} s Y_{\rm DM}\,,
\end{equation}
and satisfies $\Gamma_{\rm ann}<H|_{T_{\rm fo}}$ for all the points displayed in Fig.~\ref{fig:Therm}.}. Furthermore, as $m_V > 2 m_\pi$, the semi-annihilation channel $\pi \pi \to  \pi V$ is kinematically closed.

The Boltzmann equations for the dark meson abundances are solved for a representative benchmark point, indicated by a star in Fig.~\ref{fig:Therm}, with the corresponding results shown in the left panel of Fig.~\ref{fig:Halos}.
Although the behavior  observed in the left panel of Fig.~\ref{fig:Halos} is similar to what has been studied in the literature for standard SIMP scenarios~\cite{Hochberg:2015vrg,Katz:2020ywn,Braat:2023fhn}, our value of $m_{\pi^0}/f_\pi\sim0.6$ lies well within the perturbative regime of chiral expansion. This opens up parameter space in agreement with indirect detection constraints.   
In the numerical solution of the Boltzmann equations, we take into account the kinetic decoupling discussed in the previous section as well as the correction to the charged pion masses, which gives $\delta_V\sim 0.05$.
Such corrections have a significant impact on the DM relic density. In particular, the abundance of the charged pions (and therefore the total DM number density) at $x_{\rm fo}$ is reduced compared to the case without a portal. 
This effect, together with the kinetic decoupling, explains the fact that this benchmark point --which belongs to the vertical family of solutions discussed in Section~\ref{sec:be}-- does not reproduce the DM relic abundance for the value of $r_{12}$ along the yellow vertical lines of Fig.~\ref{fig:thetar12}, but for a slightly larger one. Nevertheless, this effect on the relic density is at most at the tens-of-percent level.

\paragraph{Constraints on DM decays.} \label{sec:constraints}
Current bounds on MeV DM  decaying to electrons require a lifetime $\tau_{\rm DM} \gtrsim 10^{26}\mathrm{\,s}$~\cite{Cirelli:2024ssz}. 
More precisely, 
gas-rich dwarf galaxies with a low cooling rate are extremely sensitive to heat injections from DM decays. The strongest constraint in the mass range 1-15 MeV is set by observations of the \textit{Leo T} dwarf galaxy \cite{Wadekar:2021qae}.
Furthermore, a DM candidate decaying to electrons would produce an imprint on charged cosmic rays. While the solar magnetic field prevents their local observation, interstellar cosmic rays can nonetheless be detected by Voyager~\cite{Boudaud:2016mos}, which sets the most stringent bound in the mass range 15-50 MeV. 
Finally, larger masses in the sub-GeV range can be probed by X-ray telescopes, which detect photons produced via Inverse Compton scattering of background radiation off electrons originating from DM decay, with the strongest constraints provided by XMM-Newton~\cite{Koechler:2023ual}.

Thus, for the choice of DM mass of 20 MeV adopted in Fig.~\ref{fig:Therm}, the strongest constraint is set by Voyager, $\tau_{\rm DM}\gtrsim 3\times 10^{26}\mathrm{\,s}$.  This is shown in  Fig.~\ref{fig:Therm}, as the yellow exclusion region. Notice that for the point indicated with the star, the DM lifetime is $\tau_{\rm DM}\simeq 5\times 10^{26}\mathrm{\,s}$.  Concerning future prospects, HERA's future measurements of the 21-cm power spectrum will probe DM lifetimes, $\tau_{\rm DM} \lesssim 10^{28}\mathrm{\,s}$~\cite{Sun:2023acy}, potentially exploring most of the parameter space in Fig.~\ref{fig:Therm}. We show this as a dotted orange line in the figure.

It is important to mention that all these bounds are obtained assuming that the DM particle decays to an $e^+ e^-$ pair. In our case, since the DM decays into $e^+e^-e^+e^-$, a correction comparable in magnitude to the lifetime itself is expected~\cite{Slatyer:2016qyl}. Nevertheless, given the large exponent of the DM mass in Eq.~\eqref{eq:DMlifetime}, this translates into a negligible correction to the DM mass and it is therefore inconsequential for Fig.~\ref{fig:Therm}. 

Finally, the $\eta$ meson can be  produced as a $s$-channel resonance from DM scatterings in galactic halos, as we discuss in more detail in Section~\ref{sec:halos}. If its branching ratio to SM particles is large enough, severe constraints apply from DM indirect searches, see e.g. Ref.~\cite{Chu:2018fzy}. 
However, the decay rate of the $\eta$ meson is dominated by the $\theta$-induced process $\eta \to 2\pi^{0}$, 
while the branching ratio into electrons is extremely suppressed, yielding no relevant experimental constraint. 
For the benchmark point shown in Fig.~\ref{fig:Halos}, 
we obtain $\text{BR}(\eta \to \text{leptons}) \sim \mathcal{O}(10^{-30})$. 
Larger values of the $\theta$ angle further reduce this branching ratio. 
Such minuscule sensitivities are far beyond the reach of any current experiment.

\paragraph{ Bounds from dark photon searches.}\label{sec:collider}
The most stringent bounds on the parameter space under consideration arise from collider, fixed-target, and beam-dump experiments. The constraints change depending on the branching ratio between visible and invisible decays. For the choice of parameters of Fig.~\ref{fig:Therm}, we have $m_V> 2m_{\eta}$ and $\alpha_d \gg \alpha_{\rm em} \epsilon$, hence the dark photon  mainly decays to the dark mesons charged under $U(1)_d$,  $V\to\pi^+ \pi^-, \, K^+ K^-$ for charge assignment 2.\footnote{
For charge assignment 1, the dark photon decays as $V \to K^+ K^-,\, K^0/\overline{K^0}$, and the corresponding constraints remain unchanged.
} Thus, the relevant limits come from searches for invisible decays.
The BaBar collaboration~\cite{BaBar:2017tiz} currently provides the most stringent bounds on invisibly decaying dark photons in the mass range of 500 MeV to 8 GeV. Future data from Belle II~\cite{Belle-II:2018jsg} are expected to significantly improve these bounds, reaching almost  $\varepsilon \simeq 10^{-4}$. 
On the other hand, among the beam dump experiments,  currently the most competitive constraint in the region relevant for thermalization below 500 MeV 
comes from NA64~\cite{NA64:2023wbi}
In the future, both NA64~\cite{NA64:2024nwj} and LDMX~\cite{LDMX:2018cma} will explore even further this parameter space, 
see e.g.~Ref.~\cite{Antel:2023hkf}. 

\section{Impact of the $\theta$ vacuum in DM halos}\label{sec:halos}

\subsection{Reconciling the relic density with  velocity-dependent scattering}

Models in which DM production occurs via $3 \to 2$ annihilations, commonly referred to as SIMP models, were initially proposed in the early 1990s~\cite{Carlson:1992fn}. These models experienced a resurgence of interest roughly a decade ago, as they naturally predict the observed DM relic abundance within regions of parameter space where the elastic self-interaction cross section per unit mass is of the order $\mathcal{O}(\text{cm}^2/\text{g})$. Scenarios with such interaction strengths are known as  SIDM models~\cite{Spergel:1999mh}, which attracted considerable attention due to their ability to reduce the central densities of DM halos. This effect appears to alleviate the tension between predictions from  collisionless CDM simulations~\cite{Dubinski:1991bm,Navarro:1995iw,Navarro:1996gj} and certain astrophysical observations~\cite{Dave:2000ar,Vogelsberger:2012ku,Rocha:2012jg,Peter:2012jh,Elbert:2014bma,Fry:2015rta}.

However, it has become increasingly evident that SIDM models with constant elastic cross sections face significant tension with observational data from galaxy clusters, which impose upper bounds on the self-interaction cross section~\cite{Harvey:2015hha, Sagunski:2020spe, Harvey:2018uwf, Bondarenko:2017rfu, DES:2023bzs}:
\begin{equation}
\frac{\sigma}{m} \lesssim \unit[0.5]{cm^2/g}.
\end{equation}

\begin{figure}[t]
     \centering
     \includegraphics[width=0.99\linewidth]{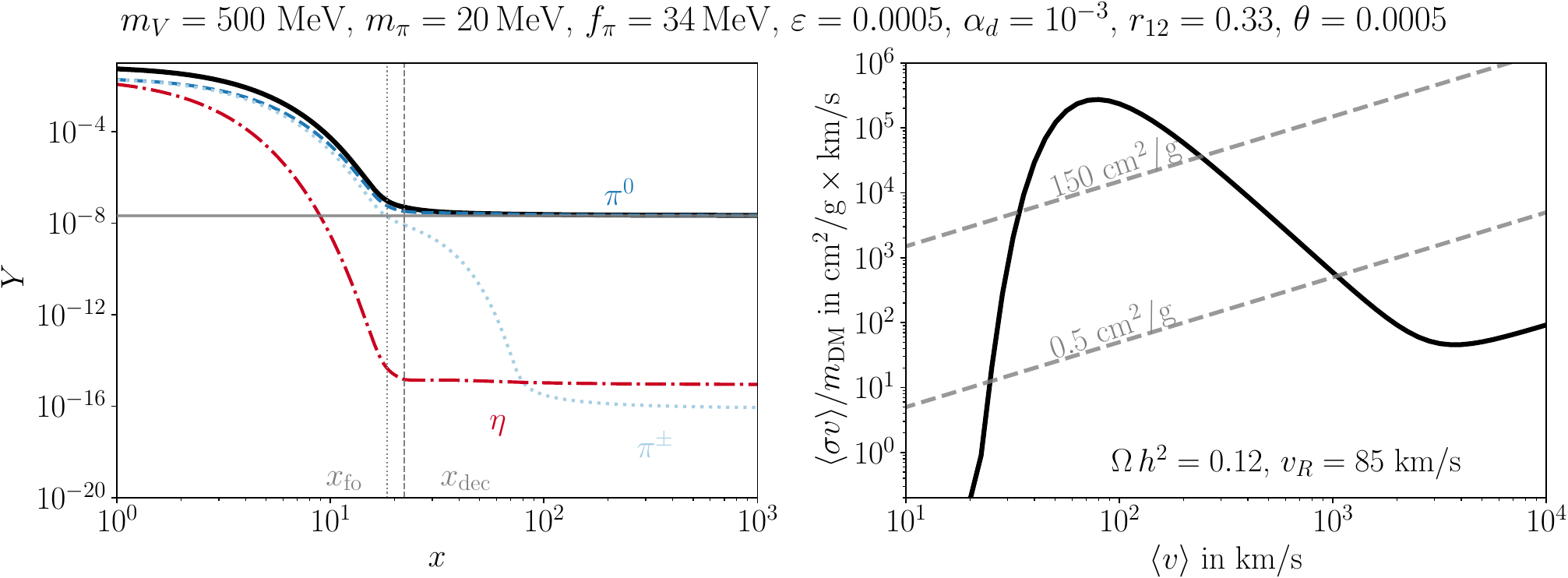}

     \caption{   
 \textbf{\textit{Left panel:}} Abundance of DM and the various species for the benchmark point indicated with a star in Figure~\ref{fig:Therm}. We use $T_{\rm dark}=T \sim 1/a$ for $x<x_{\rm dec}$ and $T_{\rm dark}=Tx_{\rm dec}/x \sim 1/a^2$ for $x > x_{\rm dec}$, as explained in Section~\ref{sec:generalT}. 
 \textbf{\textit{Right panel:}} 
Velocity dependence of the DM self-scattering cross section for the same benchmark point. For reference, we show the upper bound from cluster observations, $\sigma/m\lesssim0.5\mathrm{\,cm^{2}\,g^{-1}}$ at $\med{v}\sim2000\mathrm{\,km\,s^{-1}}$, as well as the lower bound $\sigma/m \gtrsim 150\mathrm{\,cm^{2}\,g^{-1}}$ at  $\med{v}\simeq60\mathrm{\,km\,s^{-1}}$, seemingly explaining the lensing system SDSSJ0946+1006, see text for details. }
     \label{fig:Halos}
 \end{figure}

As typical DM velocities in galaxy clusters are larger compared to those in smaller halos, viable SIDM models now frequently incorporate velocity-dependent interactions~\cite{Tulin:2017ara,Adhikari:2022sbh}. This poses a challenge for the original SIMP framework~\cite{Hochberg:2014kqa}, which, in its simplest form, does not accommodate velocity-dependent self-interactions. For example, in pion DM scenarios with $\theta = 0$, the elastic scattering cross section remains constant at non-relativistic velocities. In contrast, the $\theta$-induced DM production mechanism considered in this work naturally yields velocity-dependent self-interaction cross sections that can evade cluster constraints while remaining sizable in smaller halos. These interactions are mediated by the $\eta$ meson acting as a resonance when $\theta \neq 0$,
 and are described by the non-relativistic Breit-Wigner formula~\cite{Chu:2018fzy,Chu:2019awd}:
\begin{equation}
\label{eq:BW}
\sigma(v) = \sigma_0 + \frac{128\pi}{m_{\rm DM}^2 v_R^2} \frac{\Gamma^2}{m_{\rm DM}^2 (v^2 - v_R^2)^2 + 4 \Gamma^2 v^2 / v_R^2},
\end{equation}
where $v$ is the relative velocity of DM particles, $v_R$ is the resonant velocity defined in Eq.~\eqref{eq:vR} and $\Gamma$ is the decay rate of the mediator. For BM1 
\begin{equation}
\sigma_0 = \frac{m_{\rm DM}^2}{128\pi f_\pi^4}\,,
\end{equation}
and $\Gamma$ is given in Eq.~\eqref{eq:decay}.
In BM1, all DM today consists of $\DM$ due to the efficient conversion of $\pi^\pm$ (and heavier stable mesons) into $\DM$ in the early universe. Furthermore, we are assuming that the production of $\pi^\pm$ from $\DM$ in astrophysical halos is kinematically forbidden, which holds when $\delta \gg v^2 \sim 10^{-5}$ today, excluding the gray region shown in the left panel of Fig.~\ref{fig:r12_theta_contours}. As described in Section~\ref{sec:DPportal}, in the presence of the dark photon portal,  $\pi^\pm$  may become heavier due to radiative corrections. This makes its production in halos even harder, enlarging the viable parameter space.

Following the analysis of Ref.~\cite{Chu:2018fzy}, and for the benchmark point indicated by a star in Fig.~\ref{fig:Therm}, 
Fig.~\ref{fig:Halos} demonstrates that the velocity dependence encoded in $\sigma(v)$ enables our QCD-like DM model to realize SIDM 
while remaining consistent with both cluster constraints and the observed relic density.

\subsection{The SIDM parameter space}

The ability of SIDM to explain cored density profiles arises from the predicted elastic self-scattering between DM particles in halos, which thermalizes their velocities and enables heat transport from the outer halo to the center, thereby flattening the central density profile—a mechanism absent in standard collisionless cold DM. This thermal evolution typically proceeds through two distinct phases. 
During the \emph{core formation} phase, heat transfer from the hotter outer halo to the cooler central region leads to the development of an isothermal, constant-density core that replaces the steep central cusp predicted by collisionless models. Over longer timescales, continued energy transport causes the core to lose thermal support, transferring heat outward and initiating the \emph{core collapse} phase, also known as gravothermal collapse. In this runaway process, the core contracts while its central density rises rapidly, analogous to the gravothermal catastrophe in globular clusters. The transition between these two phases depends on not only  $\sigma(v)/m_{\rm DM}$ but also the halo mass and assembly history, and can be influenced by environmental effects such as tidal interactions. For a review with references to the original literature, we refer the reader to Ref.~\cite{Adhikari:2022sbh}.

Recent advances, largely based on non-resonant self-scattering cross sections, have identified the preferred values for $\sigma(v)/m_{\rm DM}$. In particular, successful models require strong velocity dependence, producing core formation in high-mass halos and gravothermal core collapse in low-mass halos \cite{Shah:2023qcw,Ando:2024kpk}, with outcomes influenced by assembly history and halo concentration \cite{Yang:2022mxl,Nadler:2023nrd} . Observational constraints span a broad range: as already mentioned, galaxy clusters with $v \sim 1000~\mathrm{km\,s^{-1}}$ imply $\sigma/m_{\rm DM} \lesssim 0.1$--$1~\mathrm{cm^2\,g^{-1}}$ \cite{Rocha:2012jg,Harvey:2015hha,Kaplinghat:2015aga,Sagunski:2020spe,Andrade:2020lqq}; galaxies with $v \sim 100~\mathrm{km\,s^{-1}}$ favor $\sigma/m_{\rm DM}\gtrsim 3~\mathrm{cm^2\,g^{-1}}$ \cite{Ren:2018jpt,Nadler:2023nrd,Roberts:2024uyw}; and sub-galactic systems with $v \sim 10~\mathrm{km\,s^{-1}}$ require $\sigma/m_{\rm DM} \gtrsim 5$--$10~\mathrm{cm^2\,g^{-1}}$ as well as core collapse in some halos \cite{Valli:2017ktb,Read:2018pft,Sameie:2019zfo,Silverman:2022bhs}.  Furthermore, other analyses assuming the core formation phase \cite{Hayashi:2020syu, Kamada:2023dse} find that ultra-faint dwarf galaxies impose a stringent limit of $\sigma/m_{\rm DM} \lesssim 0.1~\mathrm{cm^2\,g^{-1}}$ at very low velocities, though uncertainties remain due to limited stellar kinematic data.

In addition to this, in the last few years there is growing interest in certain diffuse satellites and dense substructures suggesting $\sigma/m_{\rm DM} \sim 100~\mathrm{cm^2\,g^{-1}}$ or greater. This includes the  lens system SDSS\,J0946+1006~\cite{Vegetti:2009cz},  as it appears to host an unusually compact and over-concentrated dark subhalo, one of the rare dark halos smaller than a galaxy that has been directly detected by means of lensing \cite{Enzi:2024ygw}. Its steep density profile sharply deviates from the predictions of collisionless DM, prompting investigations within the SIDM framework~\cite{Minor:2020hic}, particularly in the core-collapse phase. For instance, Ref.~\cite{Tajalli:2025qjx} has analyzed this system and found the observations could be explained by an SIDM halo undergoing gravothermal collapse, provided the  self-interaction cross section is much larger than $\sigma/m_{\rm DM} \sim 100~\mathrm{cm^2\,g^{-1}}$. Another study has revisited the hypothesis of a core-collapsed SIDM halo, modeling its lensing signature via an isothermal Jeans approach. This  showed that only a very massive core-collapsed halo (mass of order $10^{11}~M_{\odot}$) can reproduce the observations,
but such a halo would be expected to host a luminous galaxy, which is not observed, posing a challenge to the SIDM interpretation \cite{Li:2025kpb}. Overall, SDSS\,J0946+1006 remains a compelling but debated candidate for observing gravothermal collapse in SIDM. 
 
Motivated by this, in Fig.~\ref{fig:Halos} we indicate a lower bound~\cite{Nadler:2023nrd, Li:2025kpb} of about $\sigma/m_{\rm DM} \simeq 150~\mathrm{cm^2\,g^{-1}}$ at intermediate velocities of order $60~\mathrm{km\,s^{-1}}$, consistent with our benchmark point. In fact, once $v_R \sim 100~\mathrm{km\,s^{-1}}$ is fixed, our QCD-like framework naturally predicts cross sections of this magnitude at $v\sim v_R$ as an unavoidable outcome of the model. Larger values of the $\theta$ angle further enhance the likelihood of producing these effects as they lead to larger cross sections. Interestingly, cross sections in this range have been proposed as a mechanism for seeding black holes, particularly in the context of SDSS\,J0946+1006, which may represent a striking phenomenological consequence of resonant SIDM in the early Universe~\cite{Jiang:2025jtr, Roberts:2024wup}.

Taken at face value, these results suggest $\sigma/m_{\rm DM} \lesssim 0.1~\mathrm{cm^2\,g^{-1}}$ at both $v \gtrsim 1000~\mathrm{km\,s^{-1}}$ and $v \ll 30~\mathrm{km\,s^{-1}}$. This pattern implies a velocity-dependent cross section with a peak near $v \sim 100~\mathrm{km\,s^{-1}}$ and a steep drop toward lower velocities, consistent with models in which DM self-scattering proceeds via a resonant intermediate state by means of a Breit-Wigner cross section~\cite{Kamada:2023dse,Chu:2018fzy}.
While Fig.~\ref{fig:Halos} is in qualitative agreement with the aforementioned effects, in this work we do not attempt a comprehensive global fit, including the relic density constraint, due to the ongoing investigations associated with resonant self-interacting DM.  For instance, at present there is no $N$-body simulation for models exhibiting the velocity dependence of the Breit-Wigner form given in Eq.~\eqref{eq:BW}. 
Moreover, while a study based on a gravothermal fluid approximation has been conducted~\cite{Kamada:2023dse}, the core-collapse phase has been examined only for models with velocity-independent cross sections or for scenarios motivated by light-mediator interactions~\cite{Tran:2024vxy,Yang:2022hkm,Yang:2022zkd,Outmezguine:2022bhq}, remaining insufficiently understood for Breit-Wigner cross sections. A detailed investigation will be presented elsewhere.

Thus, QCD-like theories with a non-zero $\theta$ parameter provide a novel mechanism for DM production through number-changing interactions, while naturally accommodating velocity-dependent self-interactions. These features allow such models to evade stringent cluster constraints and remain compatible with small-scale structure observations, potentially revitalizing the SIDM framework in light of recent astrophysical data.

\section{Conclusions}
\label{sec:conclusions}

Understanding the nature of DM remains one of the most compelling and unresolved challenges in contemporary physics. Within this context, QCD-like theories that feature dark pions as DM candidates offer a particularly attractive framework due to their structural similarity with the visible sector. A recent advance in this direction~\cite{Garcia-Cely:2024ivo} demonstrated that a non-zero topological $\theta$ angle can qualitatively alter the dynamics of dark pions, notably by inducing number-changing interactions that can drive thermal freeze-out consistent with the observed relic abundance. Moreover, the same mechanism naturally leads to resonant self-interactions that may influence the distribution of DM in galactic halos, thereby realizing the 
 SIDM paradigm in a theoretically coherent manner.

In this work, we revisited the framework introduced in Ref.~\cite{Garcia-Cely:2024ivo} and addressed several open issues, particularly those related to the connection between the dark sector and the SM. By systematically analyzing the Boltzmann equations that govern the thermal history of DM, we have shown that some of the simplifying assumptions made in the original treatment, while valid, were unnecessarily restrictive. 

To provide a concrete realization of the proposed scenario, we extended the benchmark model BM1 introduced in Ref.~\cite{Garcia-Cely:2024ivo} by introducing a dark photon mediator associated with an additional $U(1)_d$ gauge symmetry. This portal interaction not only ensures thermal contact between the dark and visible sectors but also has an impact on DM stability and indirect detection signatures. We demonstrated that the extended model remains consistent with current observational constraints, including those from indirect detection experiments and astrophysical probes.

Importantly, our analysis confirms that the strongly coupled dark sector can simultaneously account for the observed relic abundance and the small-scale structure anomalies that motivate SIDM models. The presence of resonant enhancements in the self-scattering cross section enables the required velocity dependence: enhanced interactions on galactic scales and suppressed interactions on cluster scales. 

Looking forward, our results open several avenues for future research. From a theoretical standpoint, it would be valuable to explore alternative portal interactions and gauge group structures, particularly solving the Boltzmann equations for the DM distribution in phase space. On the observational front, upcoming data from indirect detection experiments, gravitational lensing studies, and precise measurements of dwarf galaxy dynamics may offer critical tests of the velocity-dependent SIDM framework outlined in this work. 
This is particularly timely in light of systems such as SDSS\,J0946+1006, which are claimed to host an SIDM halo undergoing gravothermal collapse. A thorough assessment of this requires a detailed study of the gravothermal evolution during the core-collapse phase of DM halos, in which particles self-scatter via a Breit–Wigner resonance.

In summary, our findings demonstrate that QCD-like DM theories with a non-zero $\theta$ angle furnish a viable and predictive framework that naturally unifies thermal freeze-out with small-scale structure phenomenology. By establishing a connection between relic abundance and halo dynamics through a common underlying mechanism, these models provide a compelling DM sector.

\section*{Acknowledgments}
We are indebted to Ayuki Kamada for insightful and valuable discussions.
GL is supported by the Generalitat Valenciana APOSTD/2023 Grant No. CIAPOS/2022/ 193.
C.G.C. is supported by a Ramón y Cajal contract with Ref.~RYC2020-029248-I, the Spanish National Grant PID2022-137268NA-C55 and Generalitat Valenciana through the grant CIPROM/22/69.
L.M. is supported by the Generalitat Valenciana through the grant CIACIF/2023/91. 
O.Z. has been partially supported by Sostenibilidad-UdeA, the UdeA/CODI Grant 2022-52380, and Ministerio de Ciencias Grant CD 82315 CT ICETEX 2021-1080. O.Z. also acknowledges the support of the Fundaci\'on Carolina and the hospitality of the Instituto de Física Corpuscular UV-CSIC in Valencia, Spain.

\bibliographystyle{JHEP}
\bibliography{ref}

\providecommand{\href}[2]{#2}\begingroup\raggedright\begin{thebibliography}{100}

\bibitem{Hochberg:2014kqa}
Y.~Hochberg, E.~Kuflik, H.~Murayama, T.~Volansky and J.G.~Wacker, {{Model for
  Thermal Relic Dark Matter of Strongly Interacting Massive Particles}},
  \href{https://doi.org/10.1103/PhysRevLett.115.021301}{{Phys. Rev. Lett.}
  {\bfseries 115} (2015) 021301}
  [\href{https://arxiv.org/abs/1411.3727}{{\ttfamily 1411.3727}}].

\bibitem{Garcia-Cely:2024ivo}
C.~Garc\'\i{}a-Cely, G.~Landini and O.~Zapata, {{Dark matter in QCD-like
  theories with a theta vacuum: Cosmological and astrophysical implications}},
  \href{https://doi.org/10.1103/PhysRevD.111.063044}{{Phys. Rev. D} {\bfseries
  111} (2025) 063044} [\href{https://arxiv.org/abs/2405.10367}{{\ttfamily
  2405.10367}}].

\bibitem{Kamada:2017tsq}
A.~Kamada, H.~Kim and T.~Sekiguchi, {{Axionlike particle assisted strongly
  interacting massive particle}},
  \href{https://doi.org/10.1103/PhysRevD.96.016007}{{Phys. Rev. D} {\bfseries
  96} (2017) 016007} [\href{https://arxiv.org/abs/1704.04505}{{\ttfamily
  1704.04505}}].

\bibitem{Hochberg:2015vrg}
Y.~Hochberg, E.~Kuflik and H.~Murayama, {{SIMP Spectroscopy}},
  \href{https://doi.org/10.1007/JHEP05(2016)090}{{JHEP} {\bfseries 05} (2016)
  090} [\href{https://arxiv.org/abs/1512.07917}{{\ttfamily 1512.07917}}].

\bibitem{Kuflik:2015isi}
E.~Kuflik, M.~Perelstein, N.R.-L.~Lorier and Y.-D.~Tsai, {{Elastically
  Decoupling Dark Matter}},
  \href{https://doi.org/10.1103/PhysRevLett.116.221302}{{Phys. Rev. Lett.}
  {\bfseries 116} (2016) 221302}
  [\href{https://arxiv.org/abs/1512.04545}{{\ttfamily 1512.04545}}].

\bibitem{Bernal:2015bla}
N.~Bernal, C.~Garcia-Cely and R.~Rosenfeld, {{WIMP and SIMP Dark Matter from
  the Spontaneous Breaking of a Global Group}},
  \href{https://doi.org/10.1088/1475-7516/2015/04/012}{{JCAP} {\bfseries 04}
  (2015) 012} [\href{https://arxiv.org/abs/1501.01973}{{\ttfamily
  1501.01973}}].

\bibitem{Bernal:2015xba}
N.~Bernal and X.~Chu, {{$\mathbb {Z}_2$ SIMP Dark Matter}},
  \href{https://doi.org/10.1088/1475-7516/2016/01/006}{{JCAP} {\bfseries 01}
  (2016) 006} [\href{https://arxiv.org/abs/1510.08527}{{\ttfamily
  1510.08527}}].

\bibitem{Bernal:2015ova}
N.~Bernal, X.~Chu, C.~Garcia-Cely, T.~Hambye and B.~Zaldivar, {{Production
  Regimes for Self-Interacting Dark Matter}},
  \href{https://doi.org/10.1088/1475-7516/2016/03/018}{{JCAP} {\bfseries 03}
  (2016) 018} [\href{https://arxiv.org/abs/1510.08063}{{\ttfamily
  1510.08063}}].

\bibitem{Choi:2015bya}
S.-M.~Choi and H.M.~Lee, {{SIMP dark matter with gauged Z$_{3}$ symmetry}},
  \href{https://doi.org/10.1007/JHEP09(2015)063}{{JHEP} {\bfseries 09} (2015)
  063} [\href{https://arxiv.org/abs/1505.00960}{{\ttfamily 1505.00960}}].

\bibitem{Choi:2016hid}
S.-M.~Choi and H.M.~Lee, {{Resonant SIMP dark matter}},
  \href{https://doi.org/10.1016/j.physletb.2016.04.055}{{Phys. Lett. B}
  {\bfseries 758} (2016) 47}
  [\href{https://arxiv.org/abs/1601.03566}{{\ttfamily 1601.03566}}].

\bibitem{Soni:2016gzf}
A.~Soni and Y.~Zhang, {{Hidden SU(N) Glueball Dark Matter}},
  \href{https://doi.org/10.1103/PhysRevD.93.115025}{{Phys. Rev. D} {\bfseries
  93} (2016) 115025} [\href{https://arxiv.org/abs/1602.00714}{{\ttfamily
  1602.00714}}].

\bibitem{Kamada:2016ois}
A.~Kamada, M.~Yamada, T.T.~Yanagida and K.~Yonekura, {{SIMP from a strong U(1)
  gauge theory with a monopole condensation}},
  \href{https://doi.org/10.1103/PhysRevD.94.055035}{{Phys. Rev. D} {\bfseries
  94} (2016) 055035} [\href{https://arxiv.org/abs/1606.01628}{{\ttfamily
  1606.01628}}].

\bibitem{Bernal:2017mqb}
N.~Bernal, X.~Chu and J.~Pradler, {{Simply split strongly interacting massive
  particles}}, \href{https://doi.org/10.1103/PhysRevD.95.115023}{{Phys. Rev. D}
  {\bfseries 95} (2017) 115023}
  [\href{https://arxiv.org/abs/1702.04906}{{\ttfamily 1702.04906}}].

\bibitem{Cline:2017tka}
J.M.~Cline, H.~Liu, T.~Slatyer and W.~Xue, {{Enabling Forbidden Dark Matter}},
  \href{https://doi.org/10.1103/PhysRevD.96.083521}{{Phys. Rev. D} {\bfseries
  96} (2017) 083521} [\href{https://arxiv.org/abs/1702.07716}{{\ttfamily
  1702.07716}}].

\bibitem{Choi:2017mkk}
S.-M.~Choi, H.M.~Lee and M.-S.~Seo, {{Cosmic abundances of SIMP dark matter}},
  \href{https://doi.org/10.1007/JHEP04(2017)154}{{JHEP} {\bfseries 04} (2017)
  154} [\href{https://arxiv.org/abs/1702.07860}{{\ttfamily 1702.07860}}].

\bibitem{Kuflik:2017iqs}
E.~Kuflik, M.~Perelstein, N.R.-L.~Lorier and Y.-D.~Tsai, {{Phenomenology of
  ELDER Dark Matter}}, \href{https://doi.org/10.1007/JHEP08(2017)078}{{JHEP}
  {\bfseries 08} (2017) 078}
  [\href{https://arxiv.org/abs/1706.05381}{{\ttfamily 1706.05381}}].

\bibitem{Heikinheimo:2018esa}
M.~Heikinheimo, K.~Tuominen and K.~Lang\ae{}ble, {{Hidden strongly interacting
  massive particles}}, \href{https://doi.org/10.1103/PhysRevD.97.095040}{{Phys.
  Rev. D} {\bfseries 97} (2018) 095040}
  [\href{https://arxiv.org/abs/1803.07518}{{\ttfamily 1803.07518}}].

\bibitem{Choi:2018iit}
S.-M.~Choi, H.M.~Lee, P.~Ko and A.~Natale, {{Resolving phenomenological
  problems with strongly-interacting-massive-particle models with dark vector
  resonances}}, \href{https://doi.org/10.1103/PhysRevD.98.015034}{{Phys. Rev.
  D} {\bfseries 98} (2018) 015034}
  [\href{https://arxiv.org/abs/1801.07726}{{\ttfamily 1801.07726}}].

\bibitem{Hochberg:2018rjs}
Y.~Hochberg, E.~Kuflik, R.~Mcgehee, H.~Murayama and K.~Schutz, {{Strongly
  interacting massive particles through the axion portal}},
  \href{https://doi.org/10.1103/PhysRevD.98.115031}{{Phys. Rev. D} {\bfseries
  98} (2018) 115031} [\href{https://arxiv.org/abs/1806.10139}{{\ttfamily
  1806.10139}}].

\bibitem{Bernal:2019uqr}
N.~Bernal, X.~Chu, S.~Kulkarni and J.~Pradler, {{Self-interacting dark matter
  without prejudice}},
  \href{https://doi.org/10.1103/PhysRevD.101.055044}{{Phys. Rev. D} {\bfseries
  101} (2020) 055044} [\href{https://arxiv.org/abs/1912.06681}{{\ttfamily
  1912.06681}}].

\bibitem{Choi:2019zeb}
S.-M.~Choi, H.M.~Lee, Y.~Mambrini and M.~Pierre, {{Vector SIMP dark matter with
  approximate custodial symmetry}},
  \href{https://doi.org/10.1007/JHEP07(2019)049}{{JHEP} {\bfseries 07} (2019)
  049} [\href{https://arxiv.org/abs/1904.04109}{{\ttfamily 1904.04109}}].

\bibitem{Katz:2020ywn}
A.~Katz, E.~Salvioni and B.~Shakya, {{Split SIMPs with Decays}},
  \href{https://doi.org/10.1007/JHEP10(2020)049}{{JHEP} {\bfseries 10} (2020)
  049} [\href{https://arxiv.org/abs/2006.15148}{{\ttfamily 2006.15148}}].

\bibitem{Smirnov:2020zwf}
J.~Smirnov and J.F.~Beacom, {{New Freezeout Mechanism for Strongly Interacting
  Dark Matter}}, \href{https://doi.org/10.1103/PhysRevLett.125.131301}{{Phys.
  Rev. Lett.} {\bfseries 125} (2020) 131301}
  [\href{https://arxiv.org/abs/2002.04038}{{\ttfamily 2002.04038}}].

\bibitem{Xing:2021pkb}
C.-Y.~Xing and S.-H.~Zhu, {{Dark Matter Freeze-Out via Catalyzed
  Annihilation}}, \href{https://doi.org/10.1103/PhysRevLett.127.061101}{{Phys.
  Rev. Lett.} {\bfseries 127} (2021) 061101}
  [\href{https://arxiv.org/abs/2102.02447}{{\ttfamily 2102.02447}}].

\bibitem{Braat:2023fhn}
P.~Braat and M.~Postma, {{SIMPly add a dark photon}},
  \href{https://doi.org/10.1007/JHEP03(2023)216}{{JHEP} {\bfseries 03} (2023)
  216} [\href{https://arxiv.org/abs/2301.04513}{{\ttfamily 2301.04513}}].

\bibitem{Bernreuther:2023kcg}
E.~Bernreuther, N.~Hemme, F.~Kahlhoefer and S.~Kulkarni, {{Dark matter relic
  density in strongly interacting dark sectors with light vector mesons}},
  \href{https://arxiv.org/abs/2311.17157}{{\ttfamily 2311.17157}}.

\bibitem{Garani:2021zrr}
R.~Garani, M.~Redi and A.~Tesi, {{Dark QCD matters}},
  \href{https://doi.org/10.1007/JHEP12(2021)139}{{JHEP} {\bfseries 12} (2021)
  139} [\href{https://arxiv.org/abs/2105.03429}{{\ttfamily 2105.03429}}].

\bibitem{Dey:2016qgf}
U.K.~Dey, T.N.~Maity and T.S.~Ray, {{Light Dark Matter through Assisted
  Annihilation}}, \href{https://doi.org/10.1088/1475-7516/2017/03/045}{{JCAP}
  {\bfseries 03} (2017) 045}
  [\href{https://arxiv.org/abs/1612.09074}{{\ttfamily 1612.09074}}].

\bibitem{Frumkin:2025iit}
R.~Frumkin, Y.~Hochberg, E.~Kuflik and B.~Vilk, {{Inelastically Decoupling Dark
  Matter}},  \href{https://arxiv.org/abs/2508.04772}{{\ttfamily 2508.04772}}.

\bibitem{Redi:2016kip}
M.~Redi, A.~Strumia, A.~Tesi and E.~Vigiani, {{Di-photon resonance and Dark
  Matter as heavy pions}},
  \href{https://doi.org/10.1007/JHEP05(2016)078}{{JHEP} {\bfseries 05} (2016)
  078} [\href{https://arxiv.org/abs/1602.07297}{{\ttfamily 1602.07297}}].

\bibitem{Draper:2018tmh}
P.~Draper, J.~Kozaczuk and J.-H.~Yu, {{Theta in new QCD-like sectors}},
  \href{https://doi.org/10.1103/PhysRevD.98.015028}{{Phys. Rev. D} {\bfseries
  98} (2018) 015028} [\href{https://arxiv.org/abs/1803.00015}{{\ttfamily
  1803.00015}}].

\bibitem{Abe:2024mwa}
T.~Abe, R.~Sato and T.~Yamanaka, {{Composite dark matter with forbidden
  annihilation}}, \href{https://doi.org/10.1007/JHEP09(2024)064}{{JHEP}
  {\bfseries 09} (2024) 064}
  [\href{https://arxiv.org/abs/2404.03963}{{\ttfamily 2404.03963}}].

\bibitem{Holdom:1985ag}
B.~Holdom, {{Two U(1)'s and Epsilon Charge Shifts}},
  \href{https://doi.org/10.1016/0370-2693(86)91377-8}{{Phys. Lett. B}
  {\bfseries 166} (1986) 196}.

\bibitem{Spergel:1999mh}
D.N.~Spergel and P.J.~Steinhardt, {{Observational evidence for selfinteracting
  cold dark matter}}, \href{https://doi.org/10.1103/PhysRevLett.84.3760}{{Phys.
  Rev. Lett.} {\bfseries 84} (2000) 3760}
  [\href{https://arxiv.org/abs/astro-ph/9909386}{{\ttfamily
  astro-ph/9909386}}].

\bibitem{Tulin:2017ara}
S.~Tulin and H.-B.~Yu, {{Dark Matter Self-interactions and Small Scale
  Structure}}, \href{https://doi.org/10.1016/j.physrep.2017.11.004}{{Phys.
  Rept.} {\bfseries 730} (2018) 1}
  [\href{https://arxiv.org/abs/1705.02358}{{\ttfamily 1705.02358}}].

\bibitem{Harvey:2015hha}
D.~Harvey, R.~Massey, T.~Kitching, A.~Taylor and E.~Tittley, {{The
  non-gravitational interactions of dark matter in colliding galaxy clusters}},
  \href{https://doi.org/10.1126/science.1261381}{{Science} {\bfseries 347}
  (2015) 1462} [\href{https://arxiv.org/abs/1503.07675}{{\ttfamily
  1503.07675}}].

\bibitem{Bondarenko:2017rfu}
K.~Bondarenko, A.~Boyarsky, T.~Bringmann and A.~Sokolenko, {{Constraining
  self-interacting dark matter with scaling laws of observed halo surface
  densities}}, \href{https://doi.org/10.1088/1475-7516/2018/04/049}{{JCAP}
  {\bfseries 04} (2018) 049}
  [\href{https://arxiv.org/abs/1712.06602}{{\ttfamily 1712.06602}}].

\bibitem{Harvey:2018uwf}
D.~Harvey, A.~Robertson, R.~Massey and I.G.~McCarthy, {{Observable tests of
  self-interacting dark matter in galaxy clusters: BCG wobbles in a constant
  density core}}, \href{https://doi.org/10.1093/mnras/stz1816}{{Mon. Not. Roy.
  Astron. Soc.} {\bfseries 488} (2019) 1572}
  [\href{https://arxiv.org/abs/1812.06981}{{\ttfamily 1812.06981}}].

\bibitem{Sagunski:2020spe}
L.~Sagunski, S.~Gad-Nasr, B.~Colquhoun, A.~Robertson and S.~Tulin,
  {{Velocity-dependent Self-interacting Dark Matter from Groups and Clusters of
  Galaxies}}, \href{https://doi.org/10.1088/1475-7516/2021/01/024}{{JCAP}
  {\bfseries 01} (2021) 024}
  [\href{https://arxiv.org/abs/2006.12515}{{\ttfamily 2006.12515}}].

\bibitem{DES:2023bzs}
{\scshape DES} collaboration, {{Examining the self-interaction of dark matter
  through central cluster galaxy offsets}},
  \href{https://doi.org/10.1093/mnras/stae442}{{Mon. Not. Roy. Astron. Soc.}
  {\bfseries 529} (2024) 52}
  [\href{https://arxiv.org/abs/2304.10128}{{\ttfamily 2304.10128}}].

\bibitem{Adhikari:2022sbh}
S.~Adhikari et~al., {{Astrophysical Tests of Dark Matter Self-Interactions}},
  \href{https://arxiv.org/abs/2207.10638}{{\ttfamily 2207.10638}}.

\bibitem{Vegetti:2009cz}
S.~Vegetti, L.V.E.~Koopmans, A.~Bolton, T.~Treu and R.~Gavazzi, {{Detection of
  a Dark Substructure through Gravitational Imaging}},
  \href{https://doi.org/10.1111/j.1365-2966.2010.16865.x}{{Mon. Not. Roy.
  Astron. Soc.} {\bfseries 408} (2010) 1969}
  [\href{https://arxiv.org/abs/0910.0760}{{\ttfamily 0910.0760}}].

\bibitem{Minor:2020hic}
Q.E.~Minor, S.~Gad-Nasr, M.~Kaplinghat and S.~Vegetti, {{An unexpected high
  concentration for the dark substructure in the gravitational lens
  SDSSJ0946+1006}}, \href{https://doi.org/10.1093/mnras/stab2247}{{Mon. Not.
  Roy. Astron. Soc.} {\bfseries 507} (2021) 1662}
  [\href{https://arxiv.org/abs/2011.10627}{{\ttfamily 2011.10627}}].

\bibitem{Bai:2013xga}
Y.~Bai and P.~Schwaller, {{Scale of dark QCD}},
  \href{https://doi.org/10.1103/PhysRevD.89.063522}{{Phys. Rev. D} {\bfseries
  89} (2014) 063522} [\href{https://arxiv.org/abs/1306.4676}{{\ttfamily
  1306.4676}}].

\bibitem{Pich:1991fq}
A.~Pich and E.~de~Rafael, {{Strong CP violation in an effective chiral
  Lagrangian approach}},
  \href{https://doi.org/10.1016/0550-3213(91)90019-T}{{Nucl. Phys. B}
  {\bfseries 367} (1991) 313}.

\bibitem{Scherer:2002tk}
S.~Scherer, {{Introduction to chiral perturbation theory}}, {{Adv. Nucl. Phys.}
  {\bfseries 27} (2003) 277}
  [\href{https://arxiv.org/abs/hep-ph/0210398}{{\ttfamily hep-ph/0210398}}].

\bibitem{Kamada:2022zwb}
A.~Kamada, S.~Kobayashi and T.~Kuwahara, {{Perturbative unitarity of strongly
  interacting massive particle models}},
  \href{https://doi.org/10.1007/JHEP02(2023)217}{{JHEP} {\bfseries 02} (2023)
  217} [\href{https://arxiv.org/abs/2210.01393}{{\ttfamily 2210.01393}}].

\bibitem{Wess:1971yu}
J.~Wess and B.~Zumino, {{Consequences of anomalous Ward identities}},
  \href{https://doi.org/10.1016/0370-2693(71)90582-X}{{Phys. Lett. B}
  {\bfseries 37} (1971) 95}.

\bibitem{Witten:1983tw}
E.~Witten, {{Global Aspects of Current Algebra}},
  \href{https://doi.org/10.1016/0550-3213(83)90063-9}{{Nucl. Phys. B}
  {\bfseries 223} (1983) 422}.

\bibitem{Chu:2018fzy}
X.~Chu, C.~Garcia-Cely and H.~Murayama, {{Velocity Dependence from Resonant
  Self-Interacting Dark Matter}},
  \href{https://doi.org/10.1103/PhysRevLett.122.071103}{{Phys. Rev. Lett.}
  {\bfseries 122} (2019) 071103}
  [\href{https://arxiv.org/abs/1810.04709}{{\ttfamily 1810.04709}}].

\bibitem{Tsai:2020vpi}
Y.-D.~Tsai, R.~McGehee and H.~Murayama, {{Resonant Self-Interacting Dark Matter
  from Dark QCD}}, \href{https://doi.org/10.1103/PhysRevLett.128.172001}{{Phys.
  Rev. Lett.} {\bfseries 128} (2022) 172001}
  [\href{https://arxiv.org/abs/2008.08608}{{\ttfamily 2008.08608}}].

\bibitem{Csaki:2022xmu}
C.~Cs{\'a}ki, A.~Gomes, Y.~Hochberg, E.~Kuflik, K.~Langhoff and H.~Murayama,
  {{Super-resonant dark matter}},
  \href{https://doi.org/10.1007/JHEP11(2022)162}{{JHEP} {\bfseries 11} (2022)
  162} [\href{https://arxiv.org/abs/2208.07882}{{\ttfamily 2208.07882}}].

\bibitem{Lee:2025lko}
T.~Lee and Y.-D.~Tsai, {{Naturally Resonant Dark Matter from Extra
  Dimensions}},  \href{https://arxiv.org/abs/2504.00076}{{\ttfamily
  2504.00076}}.

\bibitem{Chu:2024rrv}
X.~Chu, M.~Nikolic and J.~Pradler, {{Even SIMP miracles are possible}},
  \href{https://arxiv.org/abs/2401.12283}{{\ttfamily 2401.12283}}.

\bibitem{Planck:2018vyg}
{\scshape Planck} collaboration, {{Planck 2018 results. VI. Cosmological
  parameters}}, \href{https://doi.org/10.1051/0004-6361/201833910}{{Astron.
  Astrophys.} {\bfseries 641} (2020) A6}
  [\href{https://arxiv.org/abs/1807.06209}{{\ttfamily 1807.06209}}].

\bibitem{Frumkin:2021zng}
R.~Frumkin, Y.~Hochberg, E.~Kuflik and H.~Murayama, {{Thermal Dark Matter from
  Freeze-Out of Inverse Decays}},
  \href{https://doi.org/10.1103/PhysRevLett.130.121001}{{Phys. Rev. Lett.}
  {\bfseries 130} (2023) 121001}
  [\href{https://arxiv.org/abs/2111.14857}{{\ttfamily 2111.14857}}].

\bibitem{Garny:2017rxs}
M.~Garny, J.~Heisig, B.~L\"ulf and S.~Vogl, {{Coannihilation without chemical
  equilibrium}}, \href{https://doi.org/10.1103/PhysRevD.96.103521}{{Phys. Rev.
  D} {\bfseries 96} (2017) 103521}
  [\href{https://arxiv.org/abs/1705.09292}{{\ttfamily 1705.09292}}].

\bibitem{DAgnolo:2017dbv}
R.T.~D'Agnolo, D.~Pappadopulo and J.T.~Ruderman, {{Fourth Exception in the
  Calculation of Relic Abundances}},
  \href{https://doi.org/10.1103/PhysRevLett.119.061102}{{Phys. Rev. Lett.}
  {\bfseries 119} (2017) 061102}
  [\href{https://arxiv.org/abs/1705.08450}{{\ttfamily 1705.08450}}].

\bibitem{Puetter:2022ucx}
L.~Puetter, J.T.~Ruderman, E.~Salvioni and B.~Shakya, {{Bouncing dark matter}},
  \href{https://doi.org/10.1103/PhysRevD.109.023032}{{Phys. Rev. D} {\bfseries
  109} (2024) 023032} [\href{https://arxiv.org/abs/2208.08453}{{\ttfamily
  2208.08453}}].

\bibitem{Pappadopulo:2016pkp}
D.~Pappadopulo, J.T.~Ruderman and G.~Trevisan, {{Dark matter freeze-out in a
  nonrelativistic sector}},
  \href{https://doi.org/10.1103/PhysRevD.94.035005}{{Phys. Rev. D} {\bfseries
  94} (2016) 035005} [\href{https://arxiv.org/abs/1602.04219}{{\ttfamily
  1602.04219}}].

\bibitem{Carlson:1992fn}
E.D.~Carlson, M.E.~Machacek and L.J.~Hall, {{Self-interacting dark matter}},
  \href{https://doi.org/10.1086/171833}{{Astrophys. J.} {\bfseries 398} (1992)
  43}.

\bibitem{Stueckelberg:1938hvi}
E.C.G.~Stueckelberg, {{Interaction energy in electrodynamics and in the field
  theory of nuclear forces}},
  \href{https://doi.org/10.5169/seals-110852}{{Helv. Phys. Acta} {\bfseries 11}
  (1938) 225}.

\bibitem{Englert:1964et}
F.~Englert and R.~Brout, {{Broken Symmetry and the Mass of Gauge Vector
  Mesons}}, \href{https://doi.org/10.1103/PhysRevLett.13.321}{{Phys. Rev.
  Lett.} {\bfseries 13} (1964) 321}.

\bibitem{Higgs:1964pj}
P.W.~Higgs, {{Broken Symmetries and the Masses of Gauge Bosons}},
  \href{https://doi.org/10.1103/PhysRevLett.13.508}{{Phys. Rev. Lett.}
  {\bfseries 13} (1964) 508}.

\bibitem{Babu:1997st}
K.S.~Babu, C.F.~Kolda and J.~March-Russell, {{Implications of generalized Z -
  Z-prime mixing}}, \href{https://doi.org/10.1103/PhysRevD.57.6788}{{Phys. Rev.
  D} {\bfseries 57} (1998) 6788}
  [\href{https://arxiv.org/abs/hep-ph/9710441}{{\ttfamily hep-ph/9710441}}].

\bibitem{Cirelli:2024ssz}
M.~Cirelli, A.~Strumia and J.~Zupan, {{Dark Matter}},
  \href{https://arxiv.org/abs/2406.01705}{{\ttfamily 2406.01705}}.

\bibitem{Wadekar:2021qae}
D.~Wadekar and Z.~Wang, {{Strong constraints on decay and annihilation of dark
  matter from heating of gas-rich dwarf galaxies}},
  \href{https://doi.org/10.1103/PhysRevD.106.075007}{{Phys. Rev. D} {\bfseries
  106} (2022) 075007} [\href{https://arxiv.org/abs/2111.08025}{{\ttfamily
  2111.08025}}].

\bibitem{Boudaud:2016mos}
M.~Boudaud, J.~Lavalle and P.~Salati, {{Novel cosmic-ray electron and positron
  constraints on MeV dark matter particles}},
  \href{https://doi.org/10.1103/PhysRevLett.119.021103}{{Phys. Rev. Lett.}
  {\bfseries 119} (2017) 021103}
  [\href{https://arxiv.org/abs/1612.07698}{{\ttfamily 1612.07698}}].

\bibitem{Koechler:2023ual}
J.~Koechler, {{X-rays constraints on sub-GeV Dark Matter}},
  \href{https://doi.org/10.22323/1.441.0044}{{PoS} {\bfseries TAUP2023} (2024)
  044} [\href{https://arxiv.org/abs/2309.10043}{{\ttfamily 2309.10043}}].

\bibitem{Sun:2023acy}
Y.~Sun, J.W.~Foster, H.~Liu, J.B.~Mu\~noz and T.R.~Slatyer, {{Inhomogeneous
  energy injection in the 21-cm power spectrum: Sensitivity to dark matter
  decay}}, \href{https://doi.org/10.1103/PhysRevD.111.043015}{{Phys. Rev. D}
  {\bfseries 111} (2025) 043015}
  [\href{https://arxiv.org/abs/2312.11608}{{\ttfamily 2312.11608}}].

\bibitem{Slatyer:2016qyl}
T.R.~Slatyer and C.-L.~Wu, {{General Constraints on Dark Matter Decay from the
  Cosmic Microwave Background}},
  \href{https://doi.org/10.1103/PhysRevD.95.023010}{{Phys. Rev. D} {\bfseries
  95} (2017) 023010} [\href{https://arxiv.org/abs/1610.06933}{{\ttfamily
  1610.06933}}].

\bibitem{BaBar:2017tiz}
{\scshape BaBar} collaboration, {{Search for Invisible Decays of a Dark Photon
  Produced in ${e}^{+}{e}^{-}$ Collisions at BaBar}},
  \href{https://doi.org/10.1103/PhysRevLett.119.131804}{{Phys. Rev. Lett.}
  {\bfseries 119} (2017) 131804}
  [\href{https://arxiv.org/abs/1702.03327}{{\ttfamily 1702.03327}}].

\bibitem{Belle-II:2018jsg}
{\scshape Belle-II} collaboration, {{The Belle II Physics Book}},
  \href{https://doi.org/10.1093/ptep/ptz106}{{PTEP} {\bfseries 2019} (2019)
  123C01} [\href{https://arxiv.org/abs/1808.10567}{{\ttfamily 1808.10567}}].

\bibitem{NA64:2023wbi}
{\scshape NA64} collaboration, {{Search for Light Dark Matter with NA64 at
  CERN}}, \href{https://doi.org/10.1103/PhysRevLett.131.161801}{{Phys. Rev.
  Lett.} {\bfseries 131} (2023) 161801}
  [\href{https://arxiv.org/abs/2307.02404}{{\ttfamily 2307.02404}}].

\bibitem{NA64:2024nwj}
{\scshape NA64} collaboration, {{Shedding light on dark sectors with
  high-energy muons at the NA64 experiment at the CERN SPS}},
  \href{https://doi.org/10.1103/PhysRevD.110.112015}{{Phys. Rev. D} {\bfseries
  110} (2024) 112015} [\href{https://arxiv.org/abs/2409.10128}{{\ttfamily
  2409.10128}}].

\bibitem{LDMX:2018cma}
{\scshape LDMX} collaboration, {{Light Dark Matter eXperiment (LDMX)}},
  \href{https://arxiv.org/abs/1808.05219}{{\ttfamily 1808.05219}}.

\bibitem{Antel:2023hkf}
C.~Antel et~al., {{Feebly-interacting particles: FIPs 2022 Workshop Report}},
  \href{https://doi.org/10.1140/epjc/s10052-023-12168-5}{{Eur. Phys. J. C}
  {\bfseries 83} (2023) 1122}
  [\href{https://arxiv.org/abs/2305.01715}{{\ttfamily 2305.01715}}].

\bibitem{Dubinski:1991bm}
J.~Dubinski and R.G.~Carlberg, {{The Structure of cold dark matter halos}},
  \href{https://doi.org/10.1086/170451}{{Astrophys. J.} {\bfseries 378} (1991)
  496}.

\bibitem{Navarro:1995iw}
J.F.~Navarro, C.S.~Frenk and S.D.M.~White, {{The Structure of cold dark matter
  halos}}, \href{https://doi.org/10.1086/177173}{{Astrophys. J.} {\bfseries
  462} (1996) 563} [\href{https://arxiv.org/abs/astro-ph/9508025}{{\ttfamily
  astro-ph/9508025}}].

\bibitem{Navarro:1996gj}
J.F.~Navarro, C.S.~Frenk and S.D.M.~White, {{A Universal density profile from
  hierarchical clustering}}, \href{https://doi.org/10.1086/304888}{{Astrophys.
  J.} {\bfseries 490} (1997) 493}
  [\href{https://arxiv.org/abs/astro-ph/9611107}{{\ttfamily
  astro-ph/9611107}}].

\bibitem{Dave:2000ar}
R.~Dave, D.N.~Spergel, P.J.~Steinhardt and B.D.~Wandelt, {{Halo properties in
  cosmological simulations of selfinteracting cold dark matter}},
  \href{https://doi.org/10.1086/318417}{{Astrophys. J.} {\bfseries 547} (2001)
  574} [\href{https://arxiv.org/abs/astro-ph/0006218}{{\ttfamily
  astro-ph/0006218}}].

\bibitem{Vogelsberger:2012ku}
M.~Vogelsberger, J.~Zavala and A.~Loeb, {{Subhaloes in Self-Interacting
  Galactic Dark Matter Haloes}},
  \href{https://doi.org/10.1111/j.1365-2966.2012.21182.x}{{Mon. Not. Roy.
  Astron. Soc.} {\bfseries 423} (2012) 3740}
  [\href{https://arxiv.org/abs/1201.5892}{{\ttfamily 1201.5892}}].

\bibitem{Rocha:2012jg}
M.~Rocha, A.H.G.~Peter, J.S.~Bullock, M.~Kaplinghat, S.~Garrison-Kimmel,
  J.~Onorbe et~al., {{Cosmological Simulations with Self-Interacting Dark
  Matter I: Constant Density Cores and Substructure}},
  \href{https://doi.org/10.1093/mnras/sts514}{{Mon. Not. Roy. Astron. Soc.}
  {\bfseries 430} (2013) 81} [\href{https://arxiv.org/abs/1208.3025}{{\ttfamily
  1208.3025}}].

\bibitem{Peter:2012jh}
A.H.G.~Peter, M.~Rocha, J.S.~Bullock and M.~Kaplinghat, {{Cosmological
  Simulations with Self-Interacting Dark Matter II: Halo Shapes vs.
  Observations}}, \href{https://doi.org/10.1093/mnras/sts535}{{Mon. Not. Roy.
  Astron. Soc.} {\bfseries 430} (2013) 105}
  [\href{https://arxiv.org/abs/1208.3026}{{\ttfamily 1208.3026}}].

\bibitem{Elbert:2014bma}
O.D.~Elbert, J.S.~Bullock, S.~Garrison-Kimmel, M.~Rocha, J.~O\~norbe and
  A.H.G.~Peter, {{Core formation in dwarf haloes with self-interacting dark
  matter: no fine-tuning necessary}},
  \href{https://doi.org/10.1093/mnras/stv1470}{{Mon. Not. Roy. Astron. Soc.}
  {\bfseries 453} (2015) 29} [\href{https://arxiv.org/abs/1412.1477}{{\ttfamily
  1412.1477}}].

\bibitem{Fry:2015rta}
A.B.~Fry, F.~Governato, A.~Pontzen, T.~Quinn, M.~Tremmel, L.~Anderson et~al.,
  {{All about baryons: revisiting SIDM predictions at small halo masses}},
  \href{https://doi.org/10.1093/mnras/stv1330}{{Mon. Not. Roy. Astron. Soc.}
  {\bfseries 452} (2015) 1468}
  [\href{https://arxiv.org/abs/1501.00497}{{\ttfamily 1501.00497}}].

\bibitem{Chu:2019awd}
X.~Chu, C.~Garcia-Cely and H.~Murayama, {{A Practical and Consistent
  Parametrization of Dark Matter Self-Interactions}},
  \href{https://doi.org/10.1088/1475-7516/2020/06/043}{{JCAP} {\bfseries 06}
  (2020) 043} [\href{https://arxiv.org/abs/1908.06067}{{\ttfamily
  1908.06067}}].

\bibitem{Shah:2023qcw}
N.~Shah and S.~Adhikari, {{The abundance of core-collapsed subhaloes in SIDM:
  insights from structure formation in {\ensuremath{\Lambda}}CDM}},
  \href{https://doi.org/10.1093/mnras/stae833}{{Mon. Not. Roy. Astron. Soc.}
  {\bfseries 529} (2024) 4611}
  [\href{https://arxiv.org/abs/2308.16342}{{\ttfamily 2308.16342}}].

\bibitem{Ando:2024kpk}
S.~Ando, S.~Horigome, E.O.~Nadler, D.~Yang and H.-B.~Yu, {{SASHIMI-SIDM:
  semi-analytical subhalo modelling for self-interacting dark matter at
  sub-galactic scales}},
  \href{https://doi.org/10.1088/1475-7516/2025/02/053}{{JCAP} {\bfseries 02}
  (2025) 053} [\href{https://arxiv.org/abs/2403.16633}{{\ttfamily
  2403.16633}}].

\bibitem{Yang:2022mxl}
D.~Yang, E.O.~Nadler and H.-B.~Yu, {{Strong Dark Matter Self-interactions
  Diversify Halo Populations within and surrounding the Milky Way}},
  \href{https://doi.org/10.3847/1538-4357/acc73e}{{Astrophys. J.} {\bfseries
  949} (2023) 67} [\href{https://arxiv.org/abs/2211.13768}{{\ttfamily
  2211.13768}}].

\bibitem{Nadler:2023nrd}
E.O.~Nadler, D.~Yang and H.-B.~Yu, {{A Self-interacting Dark Matter Solution to
  the Extreme Diversity of Low-mass Halo Properties}},
  \href{https://doi.org/10.3847/2041-8213/ad0e09}{{Astrophys. J. Lett.}
  {\bfseries 958} (2023) L39}
  [\href{https://arxiv.org/abs/2306.01830}{{\ttfamily 2306.01830}}].

\bibitem{Kaplinghat:2015aga}
M.~Kaplinghat, S.~Tulin and H.-B.~Yu, {{Dark Matter Halos as Particle
  Colliders: Unified Solution to Small-Scale Structure Puzzles from Dwarfs to
  Clusters}}, \href{https://doi.org/10.1103/PhysRevLett.116.041302}{{Phys. Rev.
  Lett.} {\bfseries 116} (2016) 041302}
  [\href{https://arxiv.org/abs/1508.03339}{{\ttfamily 1508.03339}}].

\bibitem{Andrade:2020lqq}
K.E.~Andrade, J.~Fuson, S.~Gad-Nasr, D.~Kong, Q.~Minor, M.G.~Roberts et~al.,
  {{A stringent upper limit on dark matter self-interaction cross-section from
  cluster strong lensing}}, \href{https://doi.org/10.1093/mnras/stab3241}{{Mon.
  Not. Roy. Astron. Soc.} {\bfseries 510} (2021) 54}
  [\href{https://arxiv.org/abs/2012.06611}{{\ttfamily 2012.06611}}].

\bibitem{Ren:2018jpt}
T.~Ren, A.~Kwa, M.~Kaplinghat and H.-B.~Yu, {{Reconciling the Diversity and
  Uniformity of Galactic Rotation Curves with Self-Interacting Dark Matter}},
  \href{https://doi.org/10.1103/PhysRevX.9.031020}{{Phys. Rev. X} {\bfseries 9}
  (2019) 031020} [\href{https://arxiv.org/abs/1808.05695}{{\ttfamily
  1808.05695}}].

\bibitem{Roberts:2024uyw}
M.G.~Roberts, M.~Kaplinghat, M.~Valli and H.-B.~Yu, {{Gravothermal collapse and
  the diversity of galactic rotation curves}},
  \href{https://doi.org/10.1103/PhysRevD.111.103041}{{Phys. Rev. D} {\bfseries
  111} (2025) 103041} [\href{https://arxiv.org/abs/2407.15005}{{\ttfamily
  2407.15005}}].

\bibitem{Valli:2017ktb}
M.~Valli and H.-B.~Yu, {{Dark matter self-interactions from the internal
  dynamics of dwarf spheroidals}},
  \href{https://doi.org/10.1038/s41550-018-0560-7}{{Nature Astron.} {\bfseries
  2} (2018) 907} [\href{https://arxiv.org/abs/1711.03502}{{\ttfamily
  1711.03502}}].

\bibitem{Read:2018pft}
J.I.~Read, M.G.~Walker and P.~Steger, {{The case for a cold dark matter cusp in
  Draco}}, \href{https://doi.org/10.1093/mnras/sty2286}{{Mon. Not. Roy. Astron.
  Soc.} {\bfseries 481} (2018) 860}
  [\href{https://arxiv.org/abs/1805.06934}{{\ttfamily 1805.06934}}].

\bibitem{Sameie:2019zfo}
O.~Sameie, H.-B.~Yu, L.V.~Sales, M.~Vogelsberger and J.~Zavala,
  {{Self-Interacting Dark Matter Subhalos in the Milky Way{\textquoteright}s
  Tides}}, \href{https://doi.org/10.1103/PhysRevLett.124.141102}{{Phys. Rev.
  Lett.} {\bfseries 124} (2020) 141102}
  [\href{https://arxiv.org/abs/1904.07872}{{\ttfamily 1904.07872}}].

\bibitem{Silverman:2022bhs}
M.~Silverman, J.S.~Bullock, M.~Kaplinghat, V.H.~Robles and M.~Valli,
  {{Motivations for a large self-interacting dark matter cross-section~from
  Milky Way satellites}}, \href{https://doi.org/10.1093/mnras/stac3232}{{Mon.
  Not. Roy. Astron. Soc.} {\bfseries 518} (2022) 2418}
  [\href{https://arxiv.org/abs/2203.10104}{{\ttfamily 2203.10104}}].

\bibitem{Hayashi:2020syu}
K.~Hayashi, M.~Ibe, S.~Kobayashi, Y.~Nakayama and S.~Shirai, {{Probing dark
  matter self-interaction with ultrafaint dwarf galaxies}},
  \href{https://doi.org/10.1103/PhysRevD.103.023017}{{Phys. Rev. D} {\bfseries
  103} (2021) 023017} [\href{https://arxiv.org/abs/2008.02529}{{\ttfamily
  2008.02529}}].

\bibitem{Kamada:2023dse}
A.~Kamada and H.J.~Kim, {{Evolution of resonant self-interacting dark matter
  halos}}, \href{https://doi.org/10.1103/PhysRevD.109.063535}{{Phys. Rev. D}
  {\bfseries 109} (2024) 063535}
  [\href{https://arxiv.org/abs/2304.12621}{{\ttfamily 2304.12621}}].

\bibitem{Enzi:2024ygw}
W.J.R.~Enzi, C.M.~Krawczyk, D.J.~Ballard and T.E.~Collett, {{The
  overconcentrated dark halo in the strong lens SDSS J0946+1006 is a subhalo:
  evidence for self interacting dark matter?}},
  \href{https://arxiv.org/abs/2411.08565}{{\ttfamily 2411.08565}}.

\bibitem{Tajalli:2025qjx}
M.~Tajalli, S.~Vegetti, C.M.~O'Riordan, S.D.M.~White, C.D.~Fassnacht,
  D.M.~Powell et~al., {{SHARP -- IX. The dense, low-mass perturbers in
  B1938+666 and J0946+1006: implications for cold and self-interacting dark
  matter}},  \href{https://arxiv.org/abs/2505.07944}{{\ttfamily 2505.07944}}.

\bibitem{Li:2025kpb}
S.~Li et~al., {{The ''Little Dark Dot'': Evidence for Self-Interacting Dark
  Matter in the Strong Lens SDSSJ0946+1006?}},
  \href{https://arxiv.org/abs/2504.11800}{{\ttfamily 2504.11800}}.

\bibitem{Jiang:2025jtr}
F.~Jiang, Z.~Jia, H.~Zheng, L.C.~Ho, K.~Inayoshi, X.~Shen et~al., {{Formation
  of the Little Red Dots from the Core-collapse of Self-interacting Dark Matter
  Halos}},  \href{https://arxiv.org/abs/2503.23710}{{\ttfamily 2503.23710}}.

\bibitem{Roberts:2024wup}
M.G.~Roberts, L.~Braff, A.~Garg, S.~Profumo, T.~Jeltema and J.~O'Donnell,
  {{Early formation of supermassive black holes from the collapse of strongly
  self-interacting dark matter}},
  \href{https://doi.org/10.1088/1475-7516/2025/01/060}{{JCAP} {\bfseries 01}
  (2025) 060} [\href{https://arxiv.org/abs/2410.17480}{{\ttfamily
  2410.17480}}].

\bibitem{Tran:2024vxy}
V.~Tran, D.~Gilman, M.~Vogelsberger, X.~Shen, S.~O'Neil and X.~Zhang,
  {{Gravothermal Catastrophe in Resonant Self-interacting Dark Matter Models}},
   \href{https://arxiv.org/abs/2405.02388}{{\ttfamily 2405.02388}}.

\bibitem{Yang:2022hkm}
D.~Yang and H.-B.~Yu, {{Gravothermal evolution of dark matter halos with
  differential elastic scattering}},
  \href{https://doi.org/10.1088/1475-7516/2022/09/077}{{JCAP} {\bfseries 09}
  (2022) 077} [\href{https://arxiv.org/abs/2205.03392}{{\ttfamily
  2205.03392}}].

\bibitem{Yang:2022zkd}
S.~Yang, X.~Du, Z.C.~Zeng, A.~Benson, F.~Jiang, E.O.~Nadler et~al.,
  {{Gravothermal Solutions of SIDM Halos: Mapping from Constant to
  Velocity-dependent Cross Section}},
  \href{https://doi.org/10.3847/1538-4357/acbd49}{{Astrophys. J.} {\bfseries
  946} (2023) 47} [\href{https://arxiv.org/abs/2205.02957}{{\ttfamily
  2205.02957}}].

\bibitem{Outmezguine:2022bhq}
N.J.~Outmezguine, K.K.~Boddy, S.~Gad-Nasr, M.~Kaplinghat and L.~Sagunski,
  {{Universal gravothermal evolution of isolated self-interacting dark matter
  halos for velocity-dependent cross-sections}},
  \href{https://doi.org/10.1093/mnras/stad1705}{{Mon. Not. Roy. Astron. Soc.}
  {\bfseries 523} (2023) 4786}
  [\href{https://arxiv.org/abs/2204.06568}{{\ttfamily 2204.06568}}].

\end{thebibliography}\endgroup

\end{document}